\documentclass[conference,compsoc]{IEEEtran}
\usepackage{tikz}
\usepackage{amsmath}

\usepackage[left]{lineno}
\usepackage{mdframed}
\usepackage{soul}
\usepackage{fontawesome5}
\usepackage{booktabs} 
\usepackage{caption}
\usepackage{adjustbox}
\usepackage{subcaption}
\usepackage{relsize}
\usepackage{float}
\usepackage{url}
\usepackage{epstopdf}
\usepackage{graphicx} 
\usepackage{array} 
\usepackage{booktabs} 
\usepackage{tabularx}
\usepackage{tabularray}
\usepackage{graphics,graphicx}
\usepackage{balance}
\usepackage{array}
\usepackage{pifont}
\usepackage{multirow}
\usepackage{xcolor}
\usepackage{amsmath}
\usepackage{footmisc}
\usepackage{subcaption}
\usepackage{hyperref}
\usepackage[normalem]{ulem}
\usepackage{paralist, tabularx}
\usepackage{bbding}
\usepackage{pifont}
\usepackage{wasysym}
\usepackage{amssymb}
\usepackage[most]{tcolorbox}

\definecolor{fg}{RGB}{34,139,34}
\usepackage{cleveref}
\newcommand*\colourcheck[1]{%
  \expandafter\newcommand\csname #1check\endcsname{\textcolor{#1}{\ding{52}}}%
}
\newcommand*\colourtimes[1]{%
  \expandafter\newcommand\csname #1times\endcsname{\textcolor{#1}{\ding{56}}}%
}
\colourcheck{fg}
\colourtimes{red}

\newcommand{\todo}[1]{\textcolor{red}{TODO: #1}}

\newif\ifcomment
\commenttrue
\ifcomment
\newcommand{\sz}[1]{{\textcolor{red}{\textbf{Savvas: #1}}}}
\newcommand{\sep}[1]{\textcolor{blue}{\textbf{Sep: #1}}}
\newcommand{\ad}[1]{\textcolor{violet}{\textbf{Abhisek: #1}}}

\newcommand{\kg}[1]{\textcolor{magenta}{\textbf{Krishna: #1}}}
\newcommand{\stefan}[1]{\textcolor{teal}{\textbf{Stefan: #1}}}

\else
\newcommand{\sz}[1]{}
\newcommand{\sep}[1]{}
\newcommand{\ad}[1]{}
\newcommand{\kg}[1]{}
\fi

\newcommand{\changes}[1]{\textcolor{black}{#1}}
\newcommand{\bluecol}[1]{\textcolor{blue}{#1}}
\begin{document}

\date{}

\title{\Large \bf Setting the Course, but Forgetting to Steer: Analyzing Compliance with GDPR's Right of Access to Data by Instagram, TikTok, and YouTube}

\author{
Sai Keerthana Karnam$^{*1}$, 
Abhisek Dash$^{*2}$, 
Antariksh Das$^{1}$, 
Sepehr Mousavi$^{2}$,\\
Stefan Bechtold$^{3}$, 
Krishna P. Gummadi$^{2}$, 
Animesh Mukherjee$^{1}$,\\
Ingmar Weber$^{4}$, 
Savvas Zannettou$^{5}$\\[1ex]
$^{1}$IIT Kharagpur, India \quad
$^{2}$MPI--SWS, Germany \quad
$^{3}$ETH Zurich, Switzerland\\
$^{4}$UdS Saarbrücken, Germany \quad
$^{5}$TU Delft, Netherlands\\[1ex]
\small $^*$Equal contribution
}

\maketitle


\begin{abstract}
The GDPR's Right of Access aims to empower users with control over their personal data via Data Download Packages (DDPs). 
However, their effectiveness is often compromised by inconsistent platform implementations, questionable data reliability, and poor user comprehensibility. 
This paper conducts a comprehensive audit of DDPs from three social media platforms (TikTok, Instagram, and YouTube) to systematically assess these critical drawbacks. 
Despite offering similar services, we find that these platforms demonstrate significant inconsistencies in implementing the Right of Access, evident in varying levels of shared data. 
Critically, the failure to disclose processing purposes, retention periods, and other third-party data recipients serves as a further indicator of non-compliance.
Our reliability evaluations, using bots and user-donated data, reveal that while TikTok's DDPs offer more consistent and complete data, others exhibit notable shortcomings. 
Similarly, our assessment of comprehensibility, based on surveys with 400 participants, indicates that current DDPs substantially fall short of GDPR's standards.
To improve the comprehensibility, we propose and demonstrate a two-layered approach by: (1)~enhancing the data representation itself using stakeholder interpretations; and (2)~incorporating a user-friendly extension (\textit{Know Your Data}) for intuitive data visualization where users can control the level of transparency they prefer. 
Our findings underscore the need for clearer and non-conflicting regulatory guidance, stricter enforcement, and platform commitment to realize the goal of GDPR's Right of Access.
\end{abstract}
\noindent\rule{\columnwidth}{0.4pt}

\bluecol{The supplementary materials and code are available at \url{https://github.com/saikeerthana00/GDPR_Compliance}. The \textit{Know Your Data} tool can be accessed at \url{https://know-your-data.mpi-sws.org/}. This work has been accepted at IEEE S\&P 2026 as a full paper.}
\section{Introduction} \label{Sec: Intro}
The General Data Protection Regulation (GDPR)~\cite{EU2016GDPR} establishes legal requirements for data processing, security, and user rights.
One of these rights is the \textit{Right of Access by data subjects}, which includes provisions that allow data subjects (i.e., users) to request access to their data being processed by platforms (Article 15(3) GDPR).
Along with a copy of such data, GDPR also requires platforms to communicate the purpose of data collection, period of retention, and other recipients\footnote{As per Article 4 (9) of the GDPR, `recipient' means a natural or legal person, public authority, agency or body, to which personal data are disclosed~\cite{EU2016GDPR}.} of such data (Article 15(1) GDPR). 
Platforms have to comply with these requests within a stipulated period of one month. 
Platforms usually provide this data in the form of data download packages (DDPs)~\cite{valkenburg2024time}.

\noindent
\textbf{Implementation across similar platforms}: 
However, the GDPR does not provide standards for the structure or content of DDPs, leading to varying implementations of Article 15 across platforms~\cite{borem2024data}. 
Understandably, platforms that offer different types of services are expected to interpret and implement these GDPR provisions differently.
For instance, the DDP of a digital marketplace like Amazon is naturally distinct from that of a digital public space like TikTok. 
However, platforms \textit{offering similar services} should be able to provide comparable information to users.
For example, platforms providing short-format video streaming services should, in principle, offer similar details in their DDPs. 
Such consistent implementation across similar platforms is one of the important factors for enabling users to exercise their \textit{Right to Data Portability} (Article 20 GDPR), which allows users to transmit their data across platforms without hindrance~\cite{EU2016GDPR}.
Therefore, there is a need for systematic and comparative audit of implementations across platforms.
%
To the best of our knowledge, such a comparative study of the implementation of the GDPR's Right of Access across similar platforms is rare in the literature. 
These concerns lead to the following key unexplored research question:\\
\noindent
\textbf{RQ1:} \textit{How similarly do platforms offering similar services implement and thus comply with GDPR's Right of Access}?

\smallskip
\noindent
\textbf{Reliability of DDPs' content}: 
Irrespective of the similarity of implementations, a fundamental concern is the reliability of the content in DDPs. 
Lack of reliability of the DDPs undermines the overarching purpose of the Right of Access and severely compromises their utility towards users. 
Furthermore, unreliable DDPs can create a distorted understanding of data processed by platforms, limiting users' digital autonomy and right to digital privacy.
%
Such lack of reliability can also have ramifications for other stakeholders too. 
Researchers increasingly rely on these DDPs 
for studying the dynamic interplay between population and platforms~\cite{zannettou2024analyzing, yang2024coupling, vombatkere2024tiktok, wei2020twitter}.
To this end, incorrect or inconsistent information in platforms' DDPs risks the insights drawn from such studies being fundamentally flawed and misleading.
%
In fact, the above concern is no longer a mere theoretical one. 
A recently conducted survey among researchers who have used DDP in their research indicates that 86\% of them have reliability concerns with DDPs~\cite{valkenburg2024time}.
Their concerns are primarily centered around the following dimensions -- (a)~\textit{completeness}: ensuring all  relevant data is present in DDPs for the stipulated period, (b)~\textit{correctness}: verifying the accuracy of the information, such as timestamps and event sequences recorded in DDPs, and (c)~\textit{consistency}: assessing uniformity of DDPs across different requests for the same user or across users in different locations.
While researchers have reported these concerns, there is no prior systematic audit of the reliability of the DDPs, leading to our second research question:
\\
\noindent
\textbf{RQ2:} \textit{How reliable is the information within the DDPs provided by platforms?}

\smallskip
\noindent
\textbf{Comprehensibility of DDPs}: 
Even if DDPs are to contain accurate and complete information, their utility would be severely limited if users cannot easily understand the content presented to them i.e., if DDPs are not comprehensible for end-users.
In fact, to ensure comprehensibility of the communication, Article 12(1) of the GDPR states these DDPs should be (a)~concise, (b)~transparent, (c)~intelligible, (d)~easily accessible, and (e)~using clear and plain language~\cite{EU2016GDPR}.
However, neither the GDPR nor accompanying guidelines~\cite{EDPB2018Transparency} provide any clear technical definition of these requirements. 
Additionally, adherence to these requirements in the current implementations also remains unexplored in the research community.
The authors in~\cite{borem2024data} came closest to understanding the question of adherence, but they restricted their analyses to conciseness.
This gap leads to the third research question in this paper: \\
\noindent
\textbf{RQ3:} \textit{(a)~How comprehensible are the DDPs provided by platforms and (b)~how can their comprehensibility be improved?}

\begin{figure}[h]
    \centering
    \includegraphics[width=0.90\columnwidth]{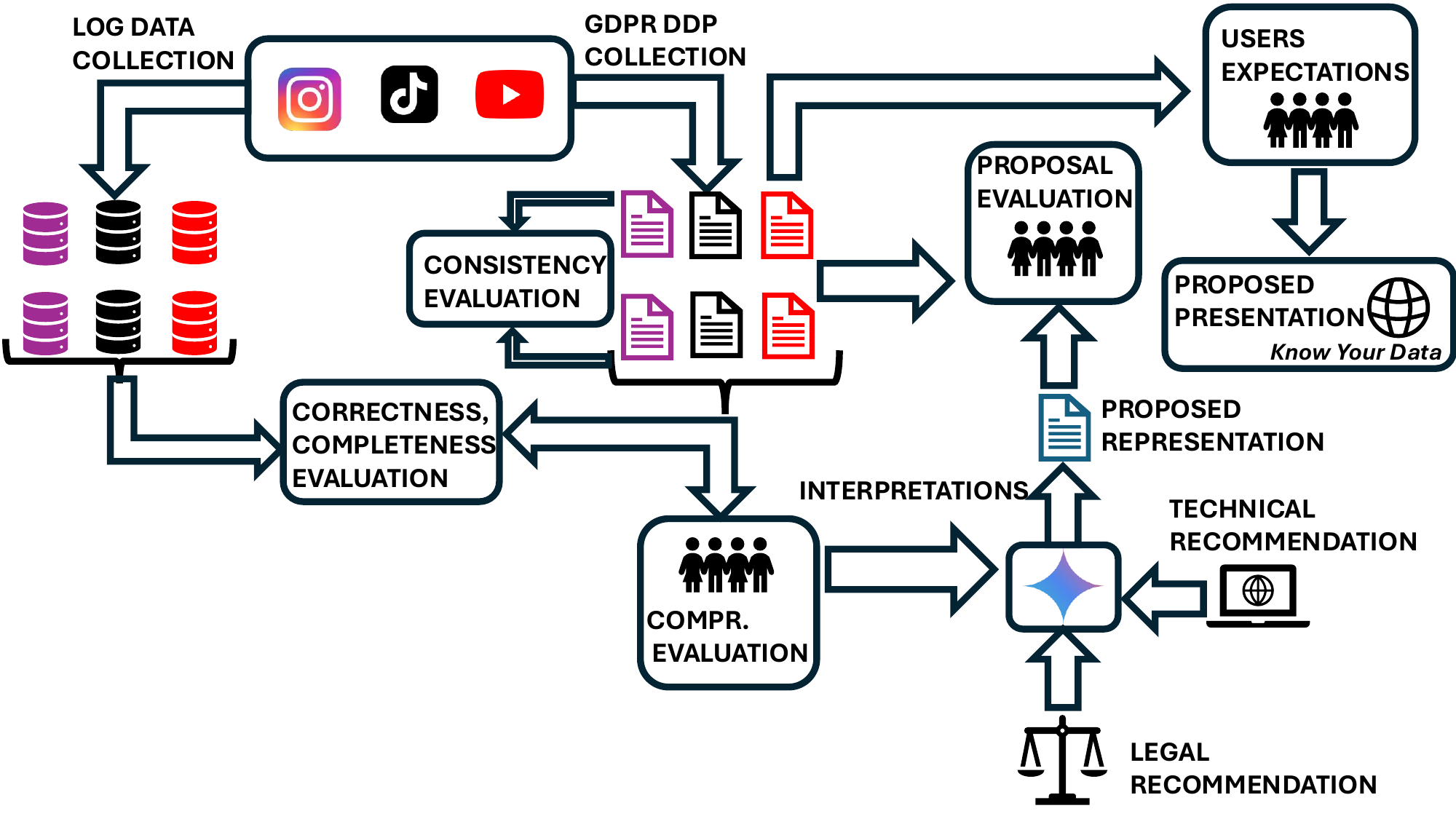}
    \vspace{-2 mm}
    \caption{Pipeline to evaluate comprehensibility and reliability of the implementation of Article 15(3) of the GDPR.}
    \vspace{-2 mm}
    \label{Fig: Pipeline}
\end{figure}

\smallskip
\noindent
\textbf{The current work}:
Although the current work could have been conducted on any digital platform, due to the growing popularity of short--format videos~\cite{zannettou2024analyzing, Mousavi_Gummadi_Zannettou_2024} and their impact~\cite{DSA2023VLOP}, we investigate the discussed research questions on TikTok, Instagram, and YouTube. \changes{\Cref{Fig: Pipeline} describes the pipeline of our conducted research.}

To audit reliability of DDPs, we browse the three platforms using sock-puppet accounts to (a)~automatically browse and log the behavior and (b)~request the corresponding DDPs from each of the platforms. 
We quantify correctness and completeness by comparing our log with the DDPs from each platform. 
We repeat this process for a period of one month to understand the consistency in the shared DDPs for the accounts. 
Also, to better understand the consistency across various accounts, we collect DDPs requested by real-world participants from the three platforms.

To assess the comprehensibility of DDPs, we conduct an extensive survey among 400 participants from Germany, France, Italy, and Spain. 
Beyond understanding the comprehensibility of the DDPs, we also ask participants about their interpretation of different GDPR prescribed requirements. 
One of the key observations in this survey is that some of the requirements are at odds with each other. 

To this end, by utilizing interpretations of participants and European Data Protection Board, we propose data representations for the different categories of information that aim at striking a better balance among these requirements.
Furthermore, to improve the comprehensibility of the DDPs, we propose a browser extension (\textit{Know Your Data}) which (a)~enables users to request and download their data in one click; (b)~provides them the autonomy to choose their desired level of transparency in data visualization. 
%
%
Below, we summarize our \textit{major findings}.

\noindent
\textbf{RQ1:} 
Instagram, TikTok, and YouTube interpret and implement GDPR's Right of Access differently, with YouTube offering surprisingly limited data compared to the others. 
Critically, none adhere to the fundamental requirement of (a)~disclosing the purpose of data collection, (b)~duration of retention, (c)~other entities with whom the data is shared for different data categories within DDPs.

\noindent
\textbf{RQ2:} 
Our analyses using sock-puppet accounts reveal TikTok's DDP presents a complete, correct, and consistent representation of our logged browsing behavior.
Instagram and YouTube exhibit varying degrees of missing data across DDPs collected during our study.
However, the analyses among real users show Instagram and TikTok share data for different periods in their DDPs for different categories, whereas YouTube shares all data for a similar duration. 

\noindent
\textbf{RQ3(a):} 
In our user survey, we find Instagram's DDP to be more comprehensible than that of the others. 
However, survey participants' interpretation of the different requirements reveals some of them may have conflicting goals; e.g., while participants interpret conciseness to prioritize relevant information, they interpret transparency to provide complete information with full disclosure.

\noindent
\textbf{RQ3(b):} 
Improving comprehensibility requires a layered approach.
One can improve the data representation themselves by accounting for interpretations from different stakeholders to achieve a better trade-off between the conflicting requirements. 
Furthermore, to improve the understandability, an additional layer of visualization in the form of a dashboard may improve the current state-of-the-art by enabling users to decide the level of conciseness/transparency they prefer.

\noindent
\textbf{Implications: }Our findings reveal several instances of platform-specific and systemic non-compliance, providing regulators with starting points to investigate the platforms.
Furthermore, insights drawn from our surveys highlight the conflicting nature of GDPR's comprehensibility requirements making it difficult for platforms to adhere to all of them at once.
At the same time, our recommendations underline the need for standardization and improved DDP communication through dashboards to ensure that the primary goals of data protection laws are achieved.

\label{app:pipeline}

\section{Background and related work}\label{Sec: Related}
\noindent
\textbf{GDPR and rights of data subjects}:
The GDPR is a landmark privacy law designed to give individuals greater control over their personal data and to standardize data protection rules across the European Union (EU)~\cite{EU2016GDPR}.
While the GDPR has many provisions and rights enlisted, our work focuses on Article 15: the \textit{Right of Access}, which allows users to access their data collected and processed by online platforms. 
Most platforms implement this requirement by providing a Data Download Package (DDP) upon request.

For such implementations, GDPR provides a set of desiderata for the DDPs.
Specifically, Article 12(1) requires a DDP to be concise, intelligible, transparent, accessible, and in clear and plain language~\cite{EU2016GDPR}.
Moreover, Article 20 gives the right to end-users to port their data from one platform to another.
However, if platforms offering similar services implement these rights in significantly different ways, platforms will essentially deprive end users of their right to portability.

\smallskip
\noindent
\textbf{Leveraging DDPs}:
Even at the current level of implementation, the richness of DDPs has enabled new research directions previously deemed infeasible.
Many recent studies are utilizing them through donations made from participants with their explicit consent.
To this extent, researchers have used DDPs to conduct research on topics related to personal health and safety~\cite{kmetty2022identifying, yang2024coupling, alsoubai2024profiling}, news and politics~\cite{blassnig2023googling, 2023haimintegratin, hase2024can}, auditing recommendation algorithms~\cite{moller2023detecting, vombatkere2024tiktok}, analyzing user behavior~\cite{zannettou2024analyzing, garimella2024whatsapp}, and ad targeting on social media platforms~\cite{wei2020twitter}.

\smallskip
\noindent
\textbf{GDPR compliance audits}:
Since the enforcement, numerous academic evaluations have been conducted to assess compliance with different provisions of GDPR. 
Prior works have investigated (a)~effective implementation of data portability~\cite{wong2018portable, syrmoudis2021data}, (b)~ethical design of consent forms~\cite{santos2019cookie, santos2021consent, matte2020cookie, bouhoula2024automated}, (c)~privacy violations under the pretense of `legitimate interest'~\cite{smith2024study, kyi2023investigating, kyi2024doesn}, and (d)~inconsistencies in platform's data withdrawal behavior~\cite{Du2024Withdraw}.

\noindent
\textbf{Current work: }However, little attention has been given to platforms' implementation of the right to access. 
While there are some studies raising privacy~\cite{borem2024data} and reliability concerns~\cite{valkenburg2024time, hase2024fulfilling}, none of the studies systematically audit the Right of Access on the legal requirements proposed by the GDPR itself.
%
%
To the best of our knowledge, this work is one of the first studies that tries to empirically evaluate the current implementations of Article 15 of the GDPR from the combined lens of comprehensibility and reliability. 
Besides auditing the compliance, we also propose a number of recommendations for policymakers and platforms alike to improve the comprehensibility 
for end-users. 

\section{Auditing compliance of similar platforms}
\label{Sec: CurrentImp}

Before analyzing reliability and comprehensibility, we first compare how platforms are currently implementing GDPR Article 15 by looking at the content of their DDPs and asses their compliance with the prescribed requirements. 
\subsection{Platforms under consideration}

\label{Sec: Platforms}

In this study, we focus on three platforms that provide \textit{short-format video} streaming services: TikTok, Instagram, and YouTube.
Our motivation to study these platforms stems from the recent popularity of short-format video platforms~\cite{zannettou2024analyzing, Mousavi_Gummadi_Zannettou_2024}.
Furthermore, all three platforms have been designated as \textit{Very Large Online Platforms}  (\textit{VLOPs}) by the European Union (EU) under the recently enacted Digital Services Act (DSA), indicating the far-reaching impact that these platforms have on the European society~\cite{DSA2023VLOP}. 

\noindent
\textbf{What do these platforms share?}: 
Upon users' request, these platforms share Data Download Packages (DDPs) primarily in two formats (a)~machine-readable -- a single or a collection of JSON/CSV file(s); (b)~human-readable -- Instagram and YouTube opt for HTML version, whereas TikTok shares in TXT files segregated across directories.

If one takes a closer look at the DDPs received from any of the platforms, their contents can be broadly divided into the following categories: (a)~user's engagement, (b)~personal details, (c)~advertisements etc.  
Next, we examine the DDPs that TikTok, Instagram, and YouTube provide and observe their degree of compliance. 

\begin{table}[t]
\centering
\resizebox{\columnwidth}{!}{
\begin{tabular}{@{}llll@{}}
\toprule
\textbf{Feature}           & \textbf{YouTube}                              & \textbf{Instagram}                                                                          & \textbf{TikTok}    \\ \midrule
\textbf{Duration}          & Entire user's lifetime                   & 2 weeks                                                                                & 6 months      \\
\textbf{Content URL}       & \textbf{\fgcheck}                             & \textbf{\redtimes}                                                                          & \textbf{\fgcheck}  \\
\textbf{Timestamp}         & \textbf{\fgcheck}                             & \textbf{\fgcheck}                                                                           & \textbf{\fgcheck}  \\
\textbf{Content title}     & \textbf{\fgcheck}                             & \textbf{\redtimes}                                                                          & \textbf{\redtimes} \\
\textbf{Author name}       & \textbf{\fgcheck} (with author's URL) & \textbf{\fgcheck}                                                                           & \textbf{\redtimes} \\
\textbf{Ad identification} & \textbf{\fgcheck} (whether ad or not)         &  \textbf{\fgcheck}                                                                          & \textbf{\redtimes} \\
\textbf{File segmentation} & Single file                                   & \begin{tabular}[c]{@{}l@{}}Three files\end{tabular} & Single file        \\ \bottomrule
\end{tabular}
}
\caption{Comparison of watch history across platforms.}
\label{tab:watch_history_comparison}
\end{table}

\begin{table}[t]
\centering
\small
\begin{tabular}{@{}lcl@{}}
\toprule
\textbf{Platform} & \textbf{\begin{tabular}[c]{@{}l@{}}Provides \end{tabular}} & \textbf{Details provided}                                                                                           \\ \midrule
TikTok            & \textbf{\fgcheck}                                                             & \begin{tabular}[c]{@{}l@{}}Link to the liked content,\\ timestamp of when it was liked\end{tabular}                \\
Instagram         & \textbf{\fgcheck}                                                             & \begin{tabular}[c]{@{}l@{}}Link to the liked content, author details,\\ timestamp of when it was liked\end{tabular} \\
YouTube           & \textbf{\redtimes}                                                             & \begin{tabular}[c]{@{}l@{}}Shows like history on mobile app and web\\version but does not share it in the DDP\end{tabular}   \\ \bottomrule
\end{tabular}
\caption{Comparison of like history across platforms.}
\label{table:like_history}
\end{table}

\begin{table*}[!h]
\small
\centering
\begin{tabular}{@{}llcccl@{}}
\toprule
\multicolumn{1}{c}{\textbf{Information}} & \multicolumn{1}{l}{\textbf{Type}} & \textbf{TikTok} & \textbf{Instagram} & \textbf{YouTube} & \multicolumn{1}{l}{\textbf{Minimum expected fields}} \\ \midrule
\multirow{10}{*}{\textbf{User's usage}}         & Watch                             & Y               & N                  & Y                & Content Id, ts                                       \\
                                         & Search                            & Y               & Y                  & Y                & Query term, ts                                       \\
                                         & Comment                           & $N^*$               & Y                  & Y                & Comment text, Content Id, ts                         \\
                                         & Like                              & Y               & Y                  & $Y^g$                & Content Id, ts                                       \\
                                         & Messages                          & Y               & Y                  & NA               & Content, User Id, ts                                 \\
                                         & Save/Favourite                    & Y               & Y                  & Y                & Content Id, ts                                       \\
                                         & Share(In-app)                     & Y               & Y                  & NA               & Content Id, ts                                       \\
                                         & Share(Across-app)                 & Y               & N                  & N                & Content Id, ts                                       \\
                                         & Interests/Topics                  & Y               & Y                  & N                & List of topics                                       \\
                                         & Time spent                        & $N^*$           & $N^*$              & $N^*$            & Duration/frequency                                   \\ \midrule
\multirow{6}{*}{\textbf{User's content}}   & Media                             & Y               & Y                  & Y                & Media file/URL                                      \\
                                         & Text details                      & Y               & Y                  & Y                & title                                                \\
                                         & Location                          & Y               & Y                  & Y                & Some place identifiers                               \\
                                         & Date time                         & Y               & Y                  & Y                & ts                                                   \\
                                         & Device                            & -               & Y                  & $Y^g$                & Device model, OS                                     \\
                                         & Other user interactions           & $N^*$           & $N^*$              & $N^*$            & Likes, Comments                                      \\ \midrule
\multirow{7}{*}{\textbf{Personal details}}       & Account details                   & Y               & Y                  & $Y^g$            & Username, DOB, Email, Profile photo                  \\
                                         & Connections                       & Y               & Y                  & Y               & Username, ts                                         \\
                                         & Login history                     & Y               & Y                  & $Y^g$            & IP, ts                                               \\
                                         & Current devices                   & Y               & Y                  & $Y^g$            & User agent                                           \\
                                         & Current camera                    & NA              & Y                  & NA               & Version/type                                         \\
                                         & Location                          & Y               & Y                  & $Y^g$            & Place identifiers                                    \\
                                         & Account changes                   & -               & Y                  & N                & Type, Old, New values, ts                            \\ \midrule
\multirow{3}{*}{\textbf{Advertisements}}   & Ads viewed                        & N               & N                  & Y                & Content Id, ts                                       \\
                                         & Personalization                    & ${N^*}$         & N                  & N                & Reasons why the ad was shown                         \\
                                         & Access to your data               & N               & Y                  & N                & Which and how (in store visit etc.)                  \\ \midrule
\multirow{3}{*}{\textbf{Miscellaneous}}          & Off-platform                      & Y               & Y                  & N                & Platform, ts, activity                               \\
                                         & Link history                      & -               & Y                  & -                & Link, ts                                             \\
                                         & Cookies                           & -               & Y                  & -                & -                                                    \\ \bottomrule
\end{tabular}
\caption{Data Transparency: Overview of shared information (as of Dec, 2024). ${N^*}$ denotes the details are found in the app but not in the GDPR dump. ${Y^g}$ denotes details are found in google's DDP, but not in YouTube's DDP. (ts: timestamp).}
\label{Tab: DDPInformation}
\end{table*}
\subsection{User engagement}
\label{Sec: Usage}

One of the most important categories of information in the DDPs is how a user engages with social media platforms. 
It may include the content that a user watches, searches, or engages with implicitly or explicitly.

\vspace{1 mm}
\noindent
\textit{Watch history}: Watch history is an ordered list of all the content that a user has watched before requesting the data.
DDPs of all three platforms contain watch history data of the data subject. 
However, as~\Cref{tab:watch_history_comparison} shows the data provided across the three platforms are widely varying. Among these, two points are strikingly different -- (a) while YouTube provides watch history for the entire \underline{lifetime} of a user on the platform, Instagram provides data for at most the last \underline{two weeks} and TikTok provides data for around \underline{six months} before the data request. These widely varying retention periods for watch history are provided without clear explanations for the criteria used, raising questions about full adherence to Article 15(1)(d) which requires information on data storage periods or the criteria for determining them. 
(b) While TikTok and YouTube provide the watch history all at once, Instagram segregates it into three files based on the type of content, i.e., ads, posts, and videos watched.
%


\vspace{1 mm}
\noindent
\textit{Like history}: Similar to watch history, like history is an ordered list of all the contents that a user has liked before requesting the data. 
\Cref{table:like_history} demonstrates that there is a clear lack of consistency in how platforms currently implement Article 15(3) GDPR.
The most surprising observation is YouTube does not even include like history in its DDP, despite `likes' clearly constituting personal data processed by the platform and thus potentially falling under the scope of data subjects' Right of Access. We would like to note that although Like history is not there in YouTube DDP, it is included in the Google DDP.

\vspace{1 mm}
\noindent
\textit{Time spent on platform}: 
Despite showing such information on their apps, \textit{none of the platforms} share the time users spend on platforms on a daily basis in the DDPs. 
Only TikTok has recently added a new field in the data, namely `Activity Summary', which enlists the number of videos a user has commented on, shared, or watched until the end. 

\noindent
\textit{Other usage activities on platform}: 
\changes{~\Cref{Tab: DDPInformation}} shows other usage activities.
Most of the activities, e.g., comment, search, save, share, and writing a message are recorded by all platforms where the features are applicable.
However, sharing across applications, i.e., when a user shares the video on other social media platforms or copies the link for posting elsewhere, is only shared in TikTok DDPs.
Also, while both TikTok and Instagram maintain a list of inferred interests,
YouTube DDPs do not have them. 


\subsection{User's content}
\label{Sec: Content}

While the usage data is about how users consume or behave on video streaming platforms, information regarding their content refers to the content that a user creates and uploads to the platform for others to consume. This information encompasses the media (image/audio/video) a user uploads on their profile, along with any textual captions, locations, date and time information, etc.
\begin{table}[t]
\centering
\resizebox{\columnwidth}{!}{
\begin{tabular}{@{}lll@{}}
\toprule
\textbf{Aspect}           & \textbf{Details provided}                                                                                                                    & \textbf{Platforms}                                                       \\ \midrule
\textbf{Content created}  & \begin{tabular}[c]{@{}l@{}}Media (image/audio/video), \\ textual captions,\\ date and time information\end{tabular}              & \begin{tabular}[c]{@{}l@{}}TikTok, \\ Instagram,\\  YouTube\end{tabular} \\ \midrule
\textbf{Addnl. details}   & \begin{tabular}[c]{@{}l@{}}Metadata such as software used\\ for uploading (e.g., Android gallery),\\ device ID, camera metadata\end{tabular} & Instagram                                                                \\ \midrule
\textbf{Location details} & \begin{tabular}[c]{@{}l@{}}Longitude and latitude of upload site\\ if the author tags the location\end{tabular}                              & \begin{tabular}[c]{@{}l@{}}TikTok,\\ Instagram,\\  YouTube\end{tabular}  \\ \bottomrule
\end{tabular}
}
\caption{User created content shared by platforms.}
\label{table:content_details}
\end{table}
While all three platforms provide a copy of the media and text details, Instagram shares additional information including software used to upload the content, device ID, metadata about the camera, etc. (\Cref{table:content_details}).

\if{0}\vspace{1 mm}
\noindent
\textit{An important omission}: While all these data are present in the DDPs currently shared by platforms, how other users on the platforms engage with a data subject's content is not shared with the data subject. \sz{why will the platform provide this though? its not about the user. i have a bit of trouble of getting the point of this paragraph and in particular failing to understand why this is important.}
From the perspective of those other users and considering their privacy concerns, the omission of these engagement data may be a justifiable action.
However, such data might be valuable if users are subjected to hate speech in response to their posts, especially when they go unnoticed by the platforms' hate speech detectors and other content moderation policies.
\stefan{I agree with Savvas. I think this is an interesting point, but it's not clear that the Right of Access entitles end users to access this information. Who watches my content on Tiktok does not involve my personal data, and so Art. 15(3) does not apply. Perhaps one could mention this as an observation, but without talking about the hate speech issues, as these are not issues regulated by the GDPR.}\fi

\subsection{Personal details}
\label{Sec: Personal}

Distinct from data on content engagement, information regarding personal details is inherently more sensitive.
These details contain basic personally identifiable information (PII), e.g., account details, one's name, phone number, e-mail address, date of birth, profile picture, etc. 
However, apart from these basic details, most of the DDPs also include a lot of other sensitive information about the users. 

\noindent
\textit{Login history}: Login history contains the list of login activities made by the data subject.
While TikTok and Instagram provide these details in their DDPs, YouTube does \textit{not} provide them in its DDP.
For YouTube, these details are usually found in the Google DDP.
The login history details are significantly different in TikTok and Instagram DDPs.
While TikTok provides the timestamp, IP, device model, operating system, network type, and carrier provider in its DDP, Instagram's login history data is significantly more detailed. 
Along with the above details, Instagram provides (and therefore collects) cookie information, language code, Instagram app version, display properties of the device, hardware identifier, and some internal identifiers.

Such differences in the amount of PII shared point to two important possibilities: (a)~TikTok might be collecting some of these details and not sharing them with its end-users and, therefore, is not being transparent; or (b)~While such detail might be argued as relevant for security purposes, the extent of Instagram's collection, particularly regarding persistent identifiers and granular device specifics, warrants scrutiny under data minimization principles (Article 5(1)(c)).
Investigating these is beyond the scope of this study, and we would like to explore these directions in future work.


\subsection{Advertisements}\label{Sec: Ads}
Users are constantly served with ads when they scroll through TikTok, Instagram, and YouTube.
However, based on our observations, there is no standardized way of including ads that users have seen in their requested DDP. 

\noindent
\textit{Ads viewed}: 
While YouTube and Instagram demarcate ads viewed by users, TikTok does not distinguish them in its DDP. 
Among the two platforms that provide ads data, Instagram maintains it in a separate list, whereas YouTube labels ads in its watch history file. 
YouTube provides ads' title, URL, advertiser details, and timestamp; in contrast, Instagram only shares the advertiser name and timestamp. 

However, neither of the three platforms provides information on ad targeting parameters.
From the ads viewed details, a user can understand these are the ads they viewed on their timeline, but why those particular ads were served to them remains an enigma.
This lack of transparency regarding ad targeting could be seen as not fully aligning with the spirit of Article 15(1)(h) concerning meaningful information about automated decision-making, including profiling.

\vspace{1 mm}
\noindent
\textit{Ad targeting data sources}: The Instagram DDP contains a list of advertisers who have used data about one's online activity or profile for targeted advertising purposes, specifically to reach them as an identified individual (as opposed to targeting based on general demographics or interests).
This list also conveys information about how these advertisers obtained users' data.
In specific, as per this information, there are three ways in which the advertiser targets a user: (a)~\textit{custom audience} -- where the advertiser targets the user through a custom audience with a list of customer data (e.g., email addresses, phone numbers, or other identifiers) which the advertiser might have obtained from other third party sources; (b)~\textit{remarketing custom audience} -- where the advertiser targets the user because (s)he might have interacted with some of the advertiser's content, visited their website, etc.; (c)~ \textit{in-person store visit} -- where advertiser can use location data or check in data because the data subject might have visited one of their physical stores. 

YouTube and TikTok DDPs do not provide any such information on advertisers and how they gather data about a particular user.
Given that ad targeting often involves processing data from various sources, this omission may be a potential violation of Article 15(1)(g) which mandates disclosing such sources and the gathered data.

\vspace{1 mm}
\noindent
\textit{Off-platform activities}: Another important source of data through which these social media platforms collect enormous amounts of digital behavioral data is off-platform activities.
This is a list of all the activities in external platforms (outside the source platform) that these platforms have tracked or linked to a user.
For instance, if a user visits a website or application that also installs one of the tracking techniques of any of the short-format platforms, then the said website can share the user's activities, such as page view, purchase, search, logins, etc., along with other activities, such as adding a product to the cart, view contents, install app, launch app, etc.
Instagram and TikTok DDPs contain off-platform activities of users, whereas YouTube does not have such data.
Both Instagram and TikTok share the external platform name, activity type, and the timestamp.

The DDPs from the three platforms also provide a variety of other details.
For example, they may contain information on link history, survey, application settings, shopping, etc.
For more details, \changes{please refer to~\Cref{Tab: DDPInformation}}.



\subsection{Key omissions in current implementation}~\label{Sec: Purpose}
While the previous sections bring out some platform specific deficiencies in the content for different categories of data, the GDPR Right of Access requires the implementation to communicate much more than just the data. 
For instance, Article 15(1) mandates the purpose of data collection, the third-party or categories of third-parties to whom such data has been disclosed, the period for which data will be retained, etc.
Worryingly, all three platforms \textit{fail} to report this information in accordance with the Right of Access. 
While we acknowledge that some of the information might be available in different policy-related pages of the platform, we posit that it is the obligation of the platform to make this information explicit and accessible. 

Furthermore, none of the DDPs provide any README or manual to help users understand the significance of the different categories of collected data. 
As explained in this section, most often, platforms provide the raw data without providing any metadata or context. 
Without proper context, digesting and making sense of such complex information may be further difficult for common users.\\ 

\noindent
\textbf{Important takeaways: }

\noindent
    \faHandPointRight~Despite offering similar services, there are significant inconsistencies in data shared (and thus gathered) by the three platforms. Instagram collects disproportionately higher granularity of personal data than TikTok and YouTube.
    
    \noindent
    \faHandPointRight~There are inconsistencies in the reporting of data within platforms. 
    For example, while some details (e.g., URL to the content) are available for one category of data (like history) they are missing in another category (watch history) for Instagram DDP.

    \noindent
    \faHandPointRight~ All the studied platforms fail to report purpose, recipients, and retention period of collected data. These shortcomings defeat the purpose of such data protection rights.
\section{Auditing reliability of DDPs}
\label{Sec: Reliability}
While the platforms differ in their data collection and sharing practices, as long as they reliably share the information with users it still is acceptable. 
Therefore, in this section, we audit the reliability of DDPs provided by the platforms along the following dimensions.\\
%
%
%
%
\noindent
\textbf{Completeness}: Evaluates whether the platforms provide a comprehensive record of all user activities or the data is subject to sampling or omissions.\\ 
\noindent
\textbf{Correctness}: Determines whether the sequence of recorded activities in the DDP accurately reflects the chronological order of events.\\
\noindent
\textbf{Consistency}: Assesses the uniformity of data within same user account upon different requests and across different user accounts.

\subsection{Data collection}
\label{Sec: Reliability_Experimental_Setup}

To evaluate completeness and correctness of the data within DDPs, we need access to the ground truth i.e., what are the actual activities done on the corresponding user account. 
Hence, we use sock-puppet accounts to systematically browse and log the behavior of the sock-puppet to generate the ground truth.  
Next, we elaborate on the detailed procedure of data collection through sock-puppet accounts.


\noindent
\textbf{Sock-puppet accounts:}
We automated video browsing and liking activities using \texttt{pyautogui}~\cite{pyautogui} to control mouse and keyboard and \texttt{opencv-python}~\cite{opencv} to detect the like button on the desktop screen.
During the activities, we captured interactions with the browser by downloading the \texttt{HAR (HTTP Archive) files} (HAR is a JSON-formatted file that logs all network traffic between the browser and the server).
We viewed between 20 to 25 short-format videos in each video browsing session on each of the three platforms.
Our bot accounts watched each video for a random duration, ranging from 15 to 60 seconds, and they randomly liked some of the videos.
This approach allowed us to gather accurate ground truth data regarding user activities. 
With this ground-truth dataset, to assess the reliability of the platform provided DDPs, we, then, exercise the GDPR Right of Access from these accounts to get the DDPs from the platforms.
This process was conducted for nearly thirty video browsing sessions between October 2024 and December 2024 for each platform to avoid any form of stochasticity in observations. 

While the above data enables us to audit the correctness and completeness of DDPs, it does not fully capture the different nuances of consistency. For example, multiple time requesting for DDPs in the above approach allows us to audit consistency within an account. 
However, we still can not claim anything about the generalizability of the observation across accounts. 
Therefore, to better assess the consistency of the DDPs, we decided to collect data from real users. 

\noindent
\textbf{Real-world users:}
To collect data from real users, we designed our own data donation website. 
To ensure user privacy, all collected data was anonymized and any PII was discarded on the front-end of our data donation website. 
During the data collection process, we intentionally mandated the donation of specific data categories for each platform.
For TikTok and YouTube, video browsing history was mandatory.
In the case of Instagram, the mandatory categories included (a) like history and (b) ads and topics, which consist of posts, ads, videos viewed, and ads clicked.
Other data categories, such as search history, comments, etc., were optional and users provided them at their discretion.
\textit{Private messages and other sensitive data were not collected}. 
\changes{We recruited the users on Prolific \cite{prolific2025prolific} and employed the platform’s screening features to select the participants who met our eligibility criteria. Specifically, participants were required to have an approval rate of at least 98\%, a minimum of 100 prior contributions and have regular engagement with Instagram, TikTok and YouTube. We also mandated that the users should have at least 90 days of activity on each of these platforms.}
Users were compensated \$5 for the mandatory categories and \$1 for any additional category. 
We collected the data from the same set of users, i.e., all the users were compensated for their data donations for all the three platforms.

\noindent
\textbf{Geo-locations for our data collection: }
First, we collected data from 10 participants from Germany, France, Italy and Spain -- countries within the European Union that have the highest GDP~\cite{statista_gdp}.
While the primary focus of our study is to assess the implementation of Right of Access under the GDPR (which affects residents within EU), to understand the potential `Brussels effect' happening in enabling Right of Access across the world, we collected data from 10 participants each from the UK, Brazil, the USA, and India\footnote{For users in India, only Instagram and YouTube data were collected, as TikTok is banned in the country.}. \Cref{Tab: Demographics} in Appendix \ref{appendix:survey} shows the user demographics. 

\subsection{Audit methodology} 

Next, we briefly mention the methods adopted to leverage the collected data to audit the reliability of the DDPs.
\noindent
$\bullet$\textbf{Completeness:} We assessed completeness by comparing the number of activities recorded in our HAR logs (ground truth) with those in the DDP for each video browsing session. 
We used the DDP that was closest in time to the corresponding HAR dump.

\noindent
$\bullet$\textbf{Correctness:} We evaluated correctness by checking whether the order of contexts was maintained and whether any arbitrary entries were present in the DDP or not. 
To achieve this, we measured the Jaccard similarity for the following aspects between HAR and DDP for each activity type:  
(1)~\textit{Date}: Consistency of recorded timestamps. 
(2)~\textit{Context}: Consistency of video IDs or author IDs. 
(3)~\textit{Overall}: A combined evaluation of date and context.

\noindent
$\bullet$\textbf{Consistency (intra-user):} Here, we determine whether a particular account receives consistent data when making multiple requests from the same platform. Towards this goal, 
we examined whether the entries present in an earlier snapshot were retained in a later snapshot. 
While for Instagram and TikTok, the snapshots were collected with a gap of 1 week, for YouTube, we used a gap of one month to avoid getting blocked by YouTube to access the requested data.

\noindent
$\bullet$\textbf{Consistency (inter-user):} While the above comparisons took place between the sock-puppet account's recorded log and DDPs, to audit inter user consistency, we use the data collected from real users. 
To assess whether all users receive a similar amount of data, we analyzed video browsing history, search history, and like history obtained from real-world users across the three platforms. 
First, we measured the duration of data that was contained in each activity history by calculating the time difference between the earliest and the latest entries in the respective activity history lists. 
Since Instagram stores related records across different files within its DDP, we defined browsing history as the combination of \textit{ads viewed}, \textit{posts viewed}, and \textit{videos watched}. 
Similarly, we also included both \textit{keyword or phrase searches} and \textit{user searches} in the search history.

\subsection{Observations}
\label{Sec: Reliability_Observations}

\noindent
\textbf{Completeness}: 
As mentioned in \Cref{Sec: Usage}, Instagram 
and TikTok do \textit{not} provide the entire watch history of a user, since they register on the platform. 
Hence, these DDPs are incomplete and \textit{potentially violate Article 15(1)(d)}. 
Next, we evaluate completeness of the provided data.

Instagram offers two video browsing feeds: (1) home feed, and (2) reels feed. 
For Instagram's home feed, all the activities (watching and liking) present in our recorded logs were also present in the requested DDPs.
However, we observed that videos watched under the \textit{reels feed} were \textit{not} present in the DDP, although like activities associated with them were recorded in the DDP. 
This observation raises serious reliability concerns on the shared DDPs of Instagram. 
On TikTok, all browsing and like activities were fully recorded, with a completeness rate of 100\%. 
For YouTube, the recorded entries accounted for 99.5\% 
of the activities observed in the HAR dump.  




\noindent
\textbf{Correctness}: 
\Cref{tab:correctness} presents the average Jaccard scores across the video browsing sessions for the three platforms.
For Instagram, there is a drop in the Jaccard scores of browsing, because of changes in usernames, i.e, account username change or account deletion.
For TikTok, Jaccard scores achieved 100\% accuracy across both browsing and like activities.
For YouTube, the average Jaccard scores are 100\% for date, 99.83\% for context, and 99.5\% overall.
The slight drop in the overall score is because of the 0.5\% missing entries discussed earlier.
For TikTok and YouTube, the recorded timestamps showed a minimal difference of $\pm$ 5 seconds compared to the HAR dump values for the same activity.
However, for Instagram, the difference was as high as up to one minute, leading to a reduction in the accuracy for date and overall. 
Thus, all the DDPs showed higher correctness during our audit. 
\begin{table}[t]
    \centering
    \small
    \begin{tabular}{@{}lrrr@{}}
    \toprule
    \textbf{Platform (Activity)} & \multicolumn{1}{c}{\textbf{Date}} & \multicolumn{1}{c}{\textbf{Context}} & \multicolumn{1}{c}{\textbf{Overall}} \\ \midrule
    Instagram (Likes)            & 100\%                             & 100\%                                & 100\%                                \\
    Instagram (Browse)           & 96\%                              & 97\%                                 & 91\%                                \\
    TikTok (Browse)     & 100\%                             & 100\%                                & 100\%                                \\
    TikTok (Likes)     & 100\%                             & 100\%                                & 100\%                                \\
    YouTube (Browse)             & 100\%                             & 99.83\%                              & 99.5\%                               \\ \bottomrule
    \end{tabular}
    \caption{Average Jaccard scores across different video browsing sessions. Concerning correctness, TikTok sits at the top, followed by YouTube and Instagram. }
    \label{tab:correctness}
\end{table}


\begin{figure*}[t]
\centering
    \begin{subfigure}[t]{0.30\textwidth}
        \centering
        \includegraphics[width=\linewidth]{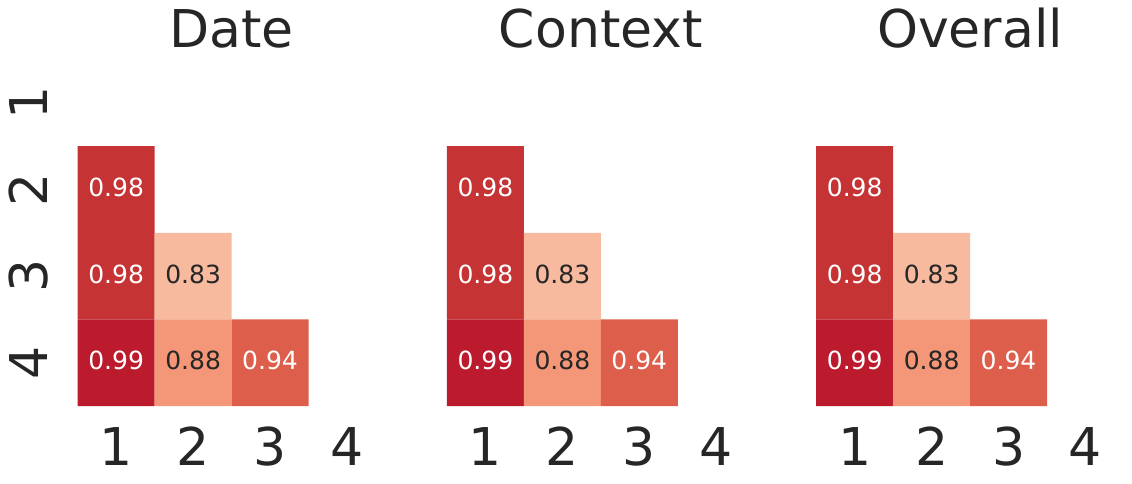}
        \caption{Instagram}
        \label{fig:IT-consistency}
    \end{subfigure}
    \hfill
    \begin{subfigure}[t]{0.30\textwidth}
        \centering
        \includegraphics[width=\linewidth]{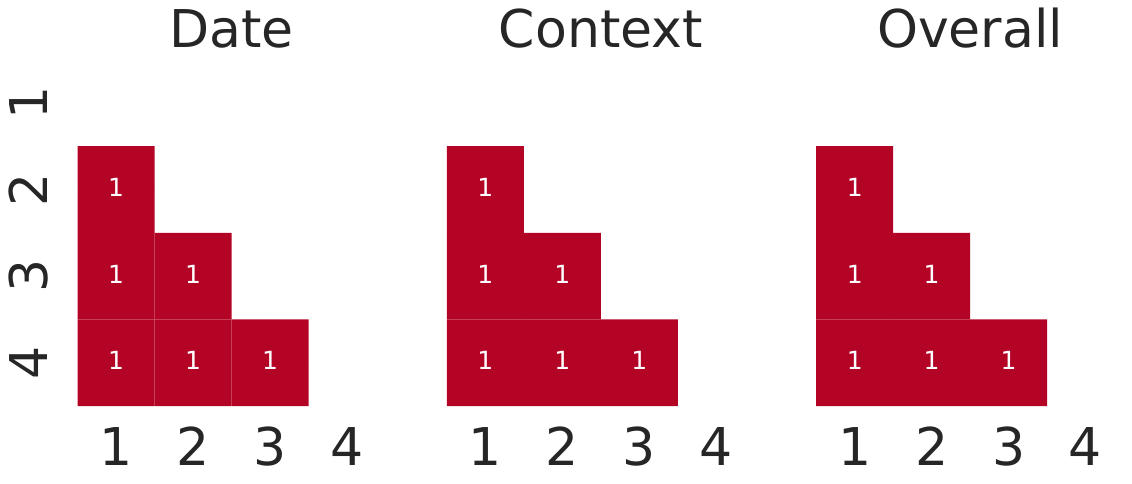}
        \caption{TikTok}
        \label{fig:TT-consistency}
    \end{subfigure}
    \hfill
    \begin{subfigure}[t]{0.30\textwidth}
        \centering
        \includegraphics[width=\linewidth]{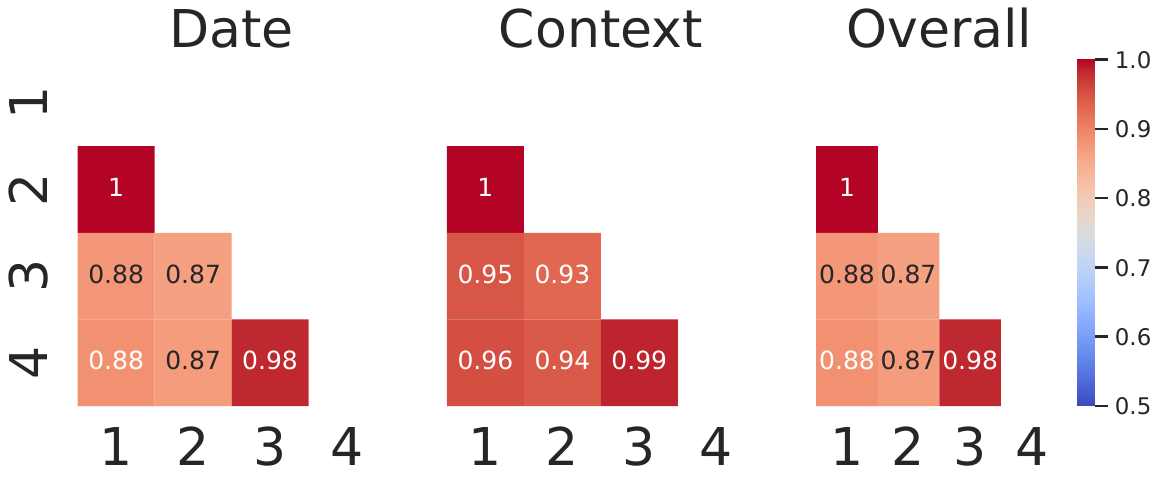}
        \caption{YouTube}
        \label{fig:YT-consistency}
    \end{subfigure}
    \caption{ Ratio of entries retained across multiple snapshots for the same account: Instagram (like history), TikTok (video browsing history and like activities), and YouTube (video browsing history).}
    \label{fig:within-consistency}
\end{figure*}


\noindent
\textbf{Consistency (intra-user)}: ~\Cref{fig:within-consistency} presents the ratio of overlapping entries between two DDPs to the total number of entries in the earlier DDP for three key aspects, i.e., date, context, and overall.
For Instagram, despite using the same accounts, we observed variations in the covered duration of the video browsing history across multiple snapshots, ranging from 6 to 13 days.
Hence, due to limited availability of browsing history, we focus on analyzing the like history.
~\Cref{fig:IT-consistency} shows the results for Instagram.
We 
observe 6\% of the entries are missing from snapshot 3 to snapshot 4.

On the other hand, ~\Cref{fig:TT-consistency} illustrates that TikTok provides consistent data across multiple retrieval attempts.
Lastly, in ~\Cref{fig:YT-consistency}, 
we observe that 12\% of entries were missing for snapshot pairs (1, 3) for YouTube.
Among these, nearly 62\% entries  were advertisements, while the other videos got deleted from the platform.
To sum up, while we found TikTok to be consistent in our audit, Instagram and YouTube failed our consistency audits. 

\begin{figure}[t] 
    \centering
    
    \begin{subfigure}[t]{0.33\columnwidth}
        \centering
        \includegraphics[width=\textwidth, height=3.0cm]{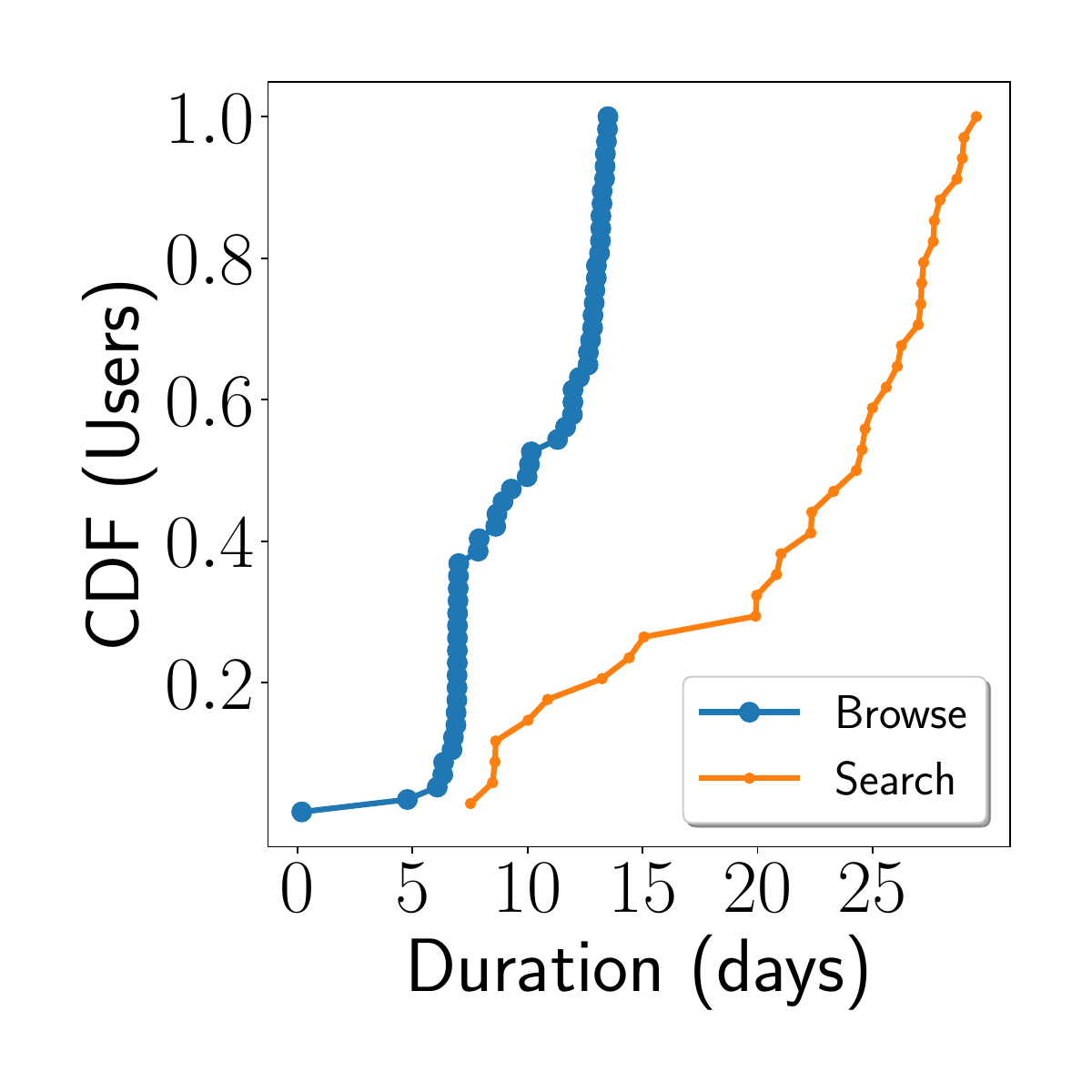}
        \caption{Instagram}
        \label{fig:insta_cdf_browse}
    \end{subfigure}
    \begin{subfigure}[t]{0.30\columnwidth}
        \centering
        \includegraphics[width=\textwidth, height=3.0cm]{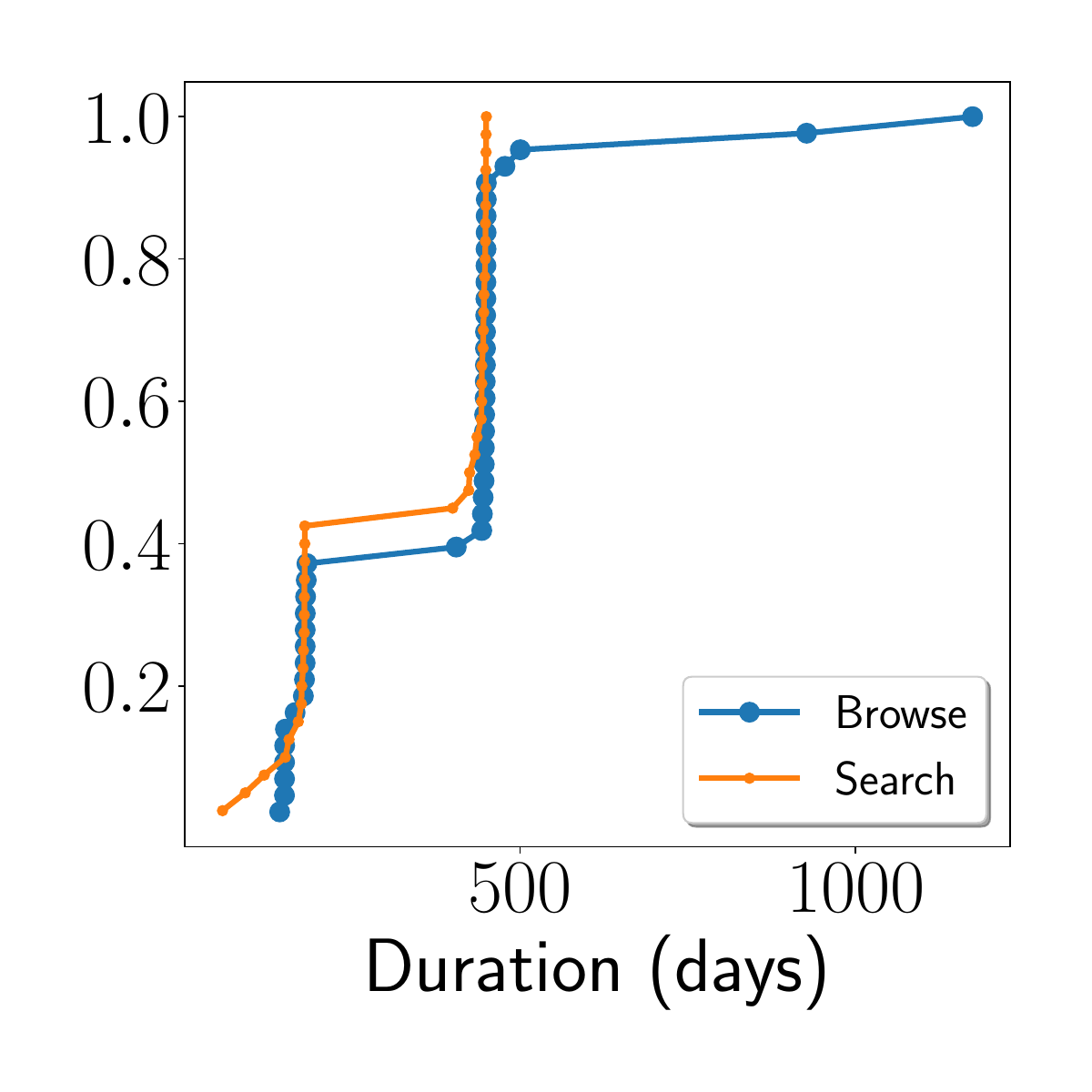}
         \caption{TikTok}
        \label{fig:tiktok_cdf_browse}
    \end{subfigure}
    \begin{subfigure}[t]{0.30\columnwidth}
        \centering
        \includegraphics[width=\textwidth, height=3.0cm]{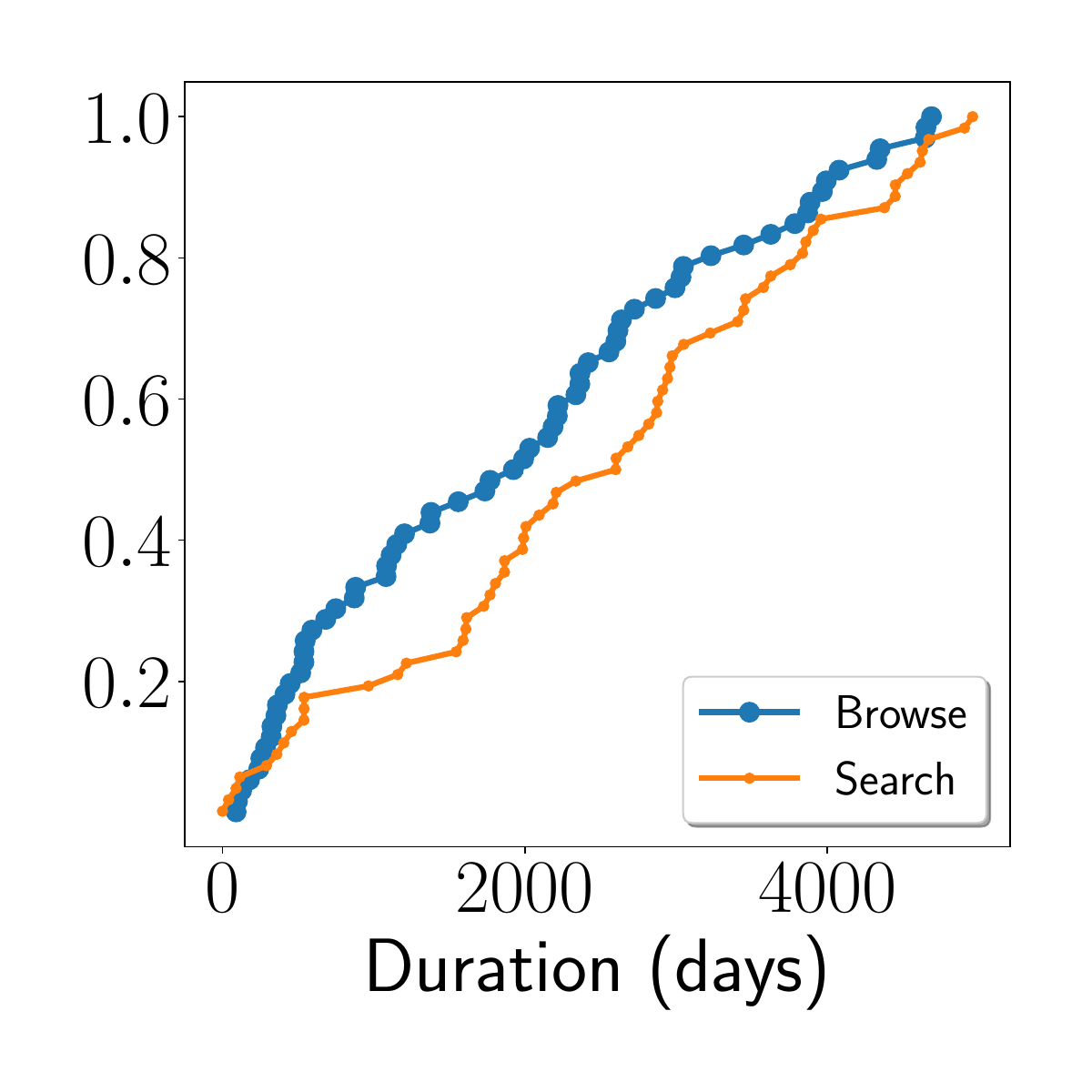}
         \caption{YouTube}
        \label{fig:youtube_cdf_browse}
    \end{subfigure}
    \hfill

    \caption{Plots illustrating the CDF of data duration provided to users for two activities -- browse and search history. On Instagram, clusters in browsing history are observed at 6 and 13 days, while on TikTok, clusters appear at 180 and 455 days for both search and browse history. 
    }
    

    \label{browse_search_overall}
\end{figure}

\noindent
\textbf{Consistency (inter-user)}:
However, these observation could be stochastic for being observed in one sock-puppet account per platform. 
To understand the extent to which the distribution of provided data differs across users, we analyzed the cumulative distribution function (CDF) of the durations of video browsing, search, and like history across the platforms. 
\Cref{browse_search_overall} presents a comparative view of activity durations across users.
For Instagram, we observe distinct clusters of users with browsing history durations concentrated around 6 and 13 days (\Cref{fig:insta_cdf_browse}). This non-uniform distribution has also been observed within the same user upon multiple requests over time.

On TikTok (\Cref{fig:tiktok_cdf_browse}), video browsing and search history durations exhibit substantial variation.
Specifically, some end-users had data spanning approximately 180 days, while others had nearly 455 days of recorded video browsing and search activities.
To ensure a fair comparison, we specifically restricted our analysis to TikTok users who have been using TikTok for at least 450 days. This criterion allows us to focus on the subset of users who have had ample time to accumulate browsing and search history.
As a result, we analyzed data from 49 TikTok users.
These findings highlight TikTok's tendency to provide data \textit{inconsistently} among its users.
However, such clusters were not observed for like history on both Instagram and TikTok. (See \Cref{cdf_ike_overall} in Appendix \ref{appendix:reliability}).
The analysis of video browsing and search history on YouTube (\Cref{fig:youtube_cdf_browse}) does not reveal any discernible patterns, suggesting that the platform may provide DDPs more consistently to its end users. This observation holds across both the EU and the other regions.

\begin{figure}[t] 
    \centering

    \begin{subfigure}[t]{0.443\columnwidth}
        \centering
        \includegraphics[width=\textwidth, height=3cm]{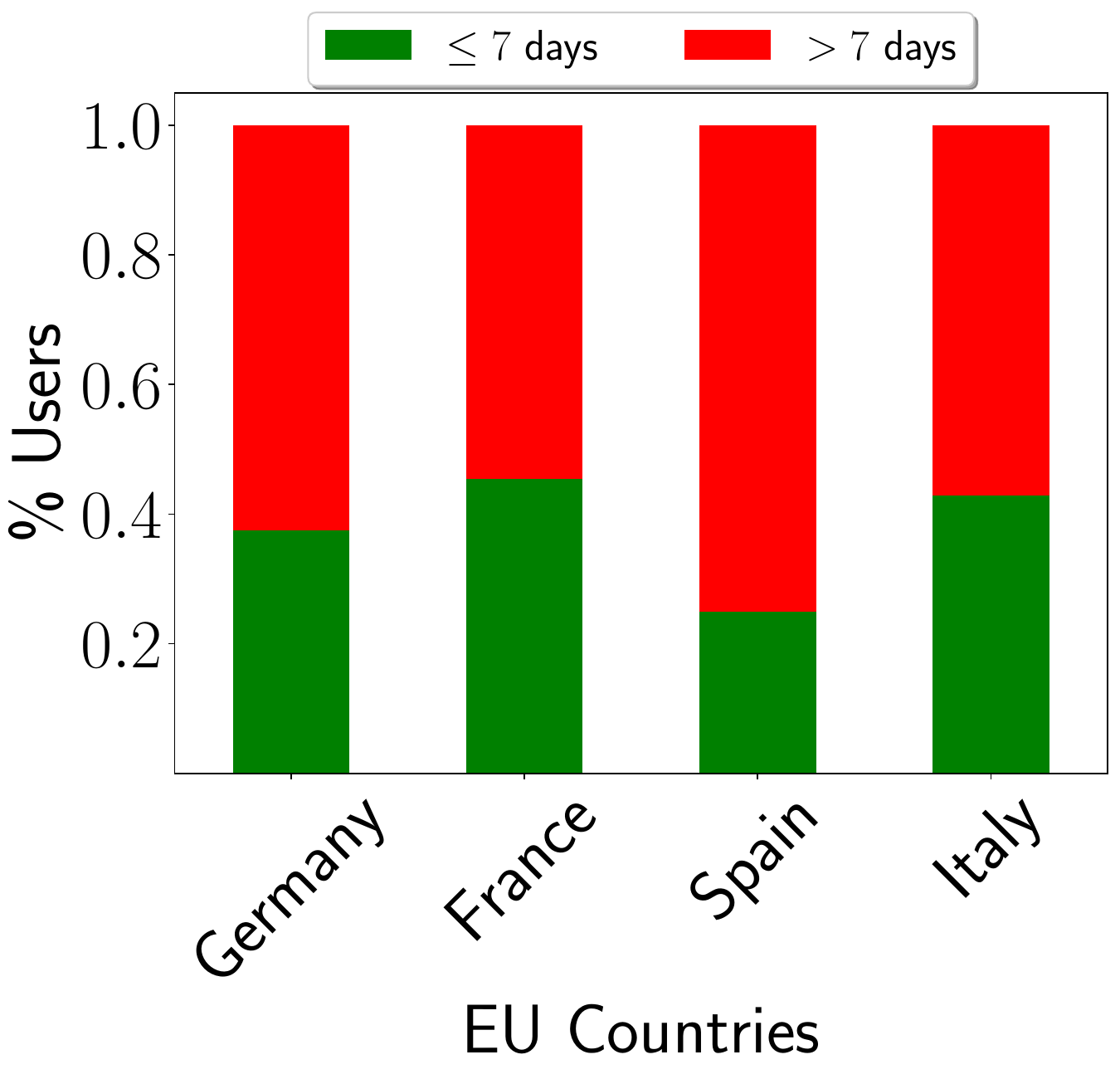}
         \caption{Instagram (EU)}
        \label{fig:instagram_cdf_eu}
    \end{subfigure}
    \begin{subfigure}[t]{0.425\columnwidth}
        \centering
        \includegraphics[width=\textwidth,height=3cm]{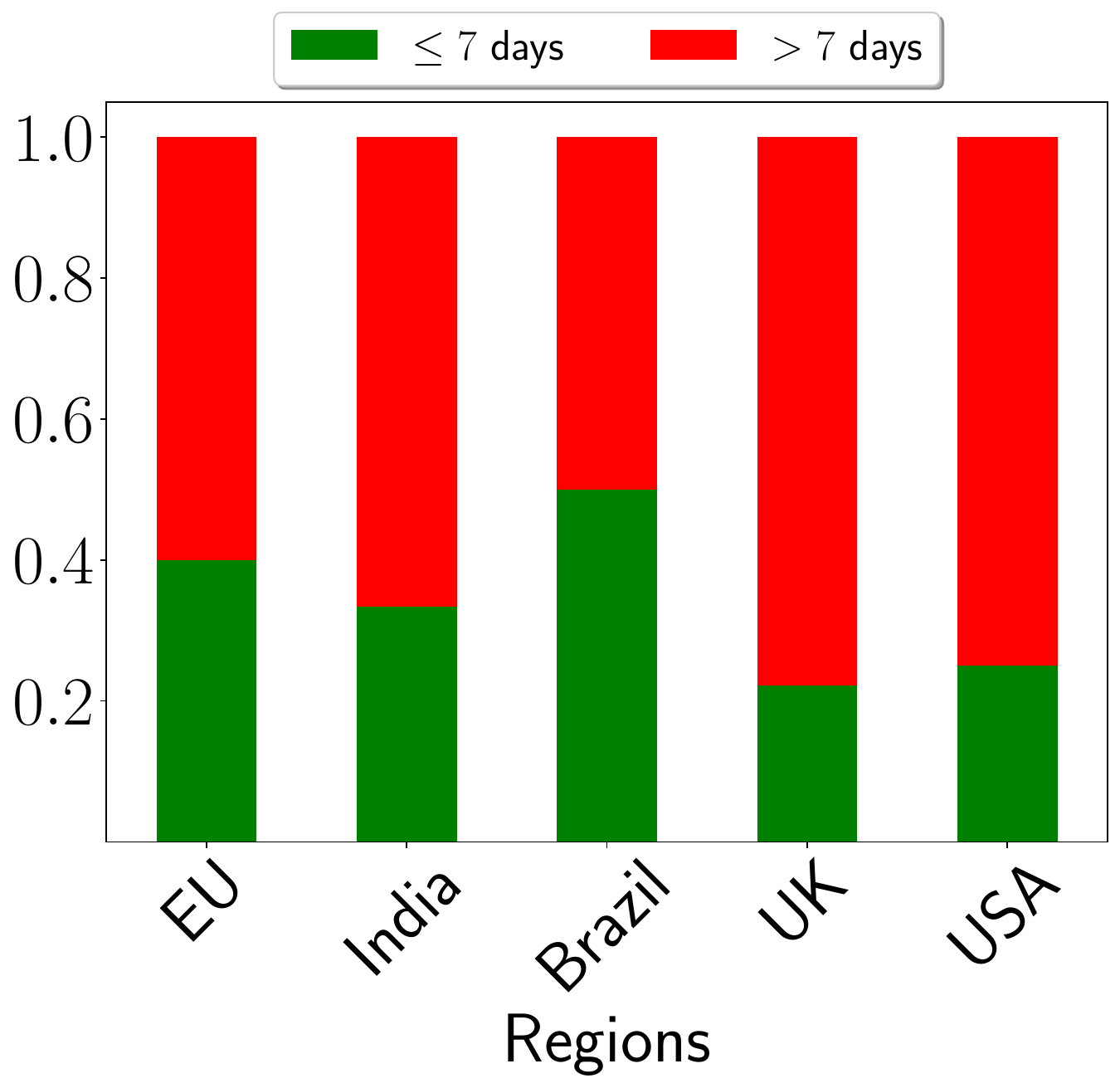}
         \caption{Instagram (Regions)}
        \label{fig:instagram_cdf_regions}
    \end{subfigure}
    \hfill
    \begin{subfigure}[t]{0.425\columnwidth}
        \centering
        \includegraphics[width=\textwidth, height=3cm]{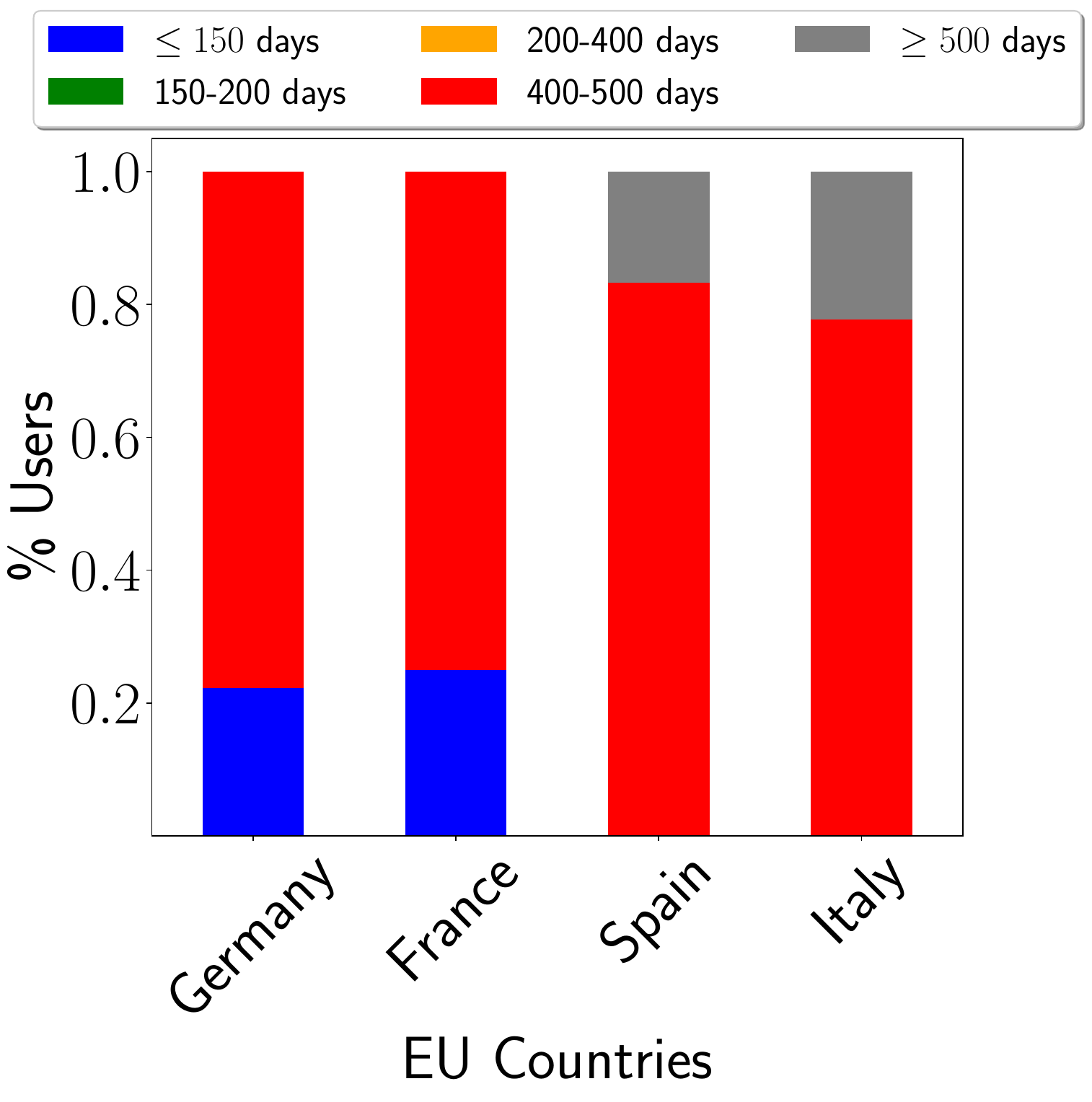}
        \caption{TikTok (EU)}
        \label{fig:tiktok_cdf_eu}
    \end{subfigure}
     \begin{subfigure}[t]{0.425\columnwidth}
        \centering
        \includegraphics[width=\textwidth, height=3cm]{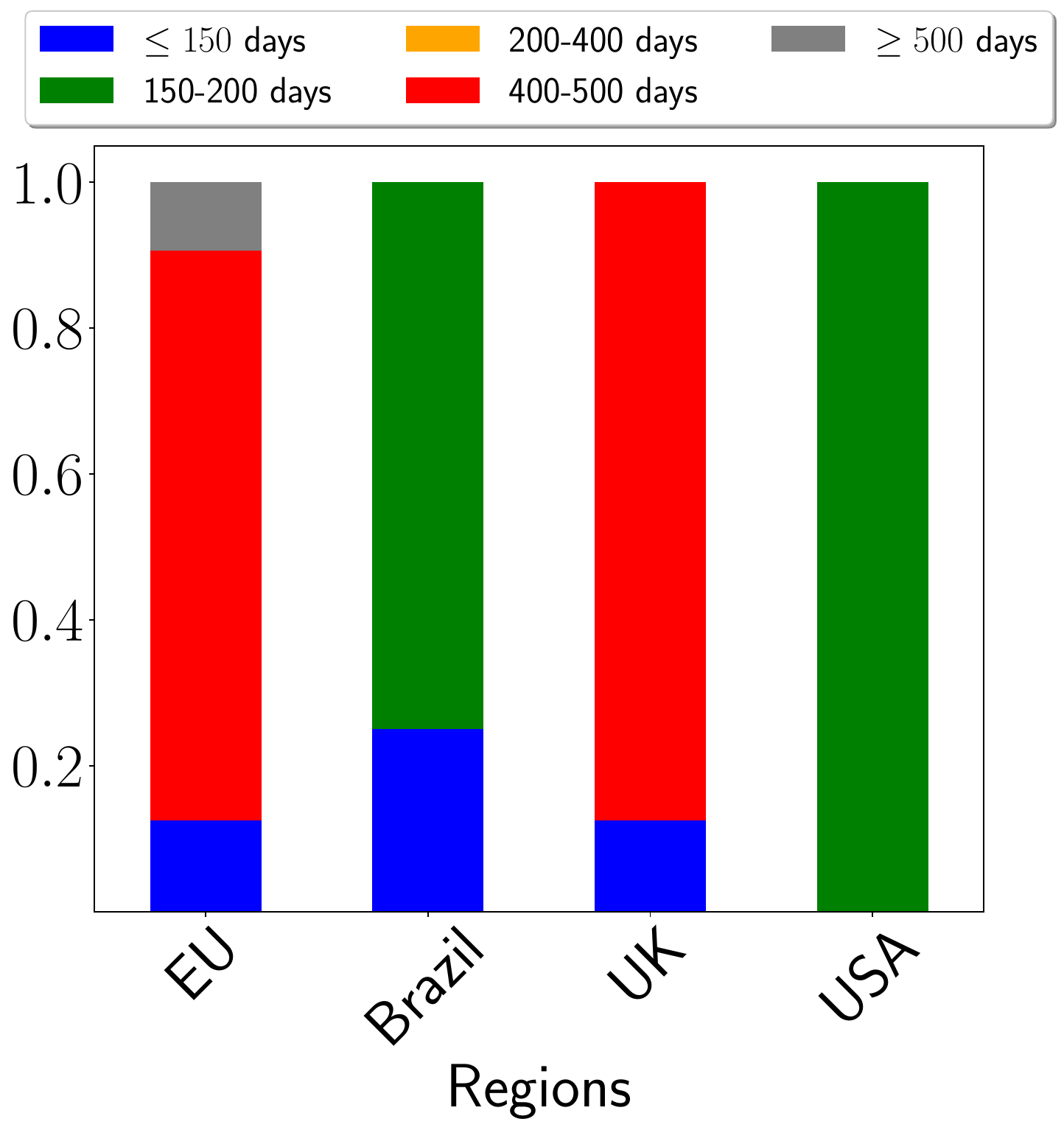}
        \caption{TikTok (Regions)}
        \label{fig:tiktok_cdf_regions}
    \end{subfigure}

    \caption{Plots showing the percentage of users with different browsing history durations. For Instagram, users within the same country or region were either given one week or two weeks of data. TikTok, by contrast, exhibits greater variability in the EU whereas in the USA, users were provided with the same duration of data.}
    

    \label{browse_across_eu_regions}
\end{figure}

\noindent
\textit{Inter-user consistency (across EU countries)}: 
Such inter-user consistency analysis shows even more interesting patterns across different countries within the EU. 
Instagram users in EU countries 
typically receive either one week or two weeks of browsing history (see~\Cref{fig:instagram_cdf_eu}). 
In contrast, TikTok displays considerable inconsistency: users within the same countries (Germany and France) often receive different duration of data, despite being part of the same regulatory region (see~\Cref{fig:tiktok_cdf_eu}). 

\noindent
\textit{Inter-user consistency (across regions)}: 
We observe the `Brussels effect' in terms of users in various other countries (UK, Brazil, USA and India) it is also possible to request and get a copy of personal data. 
Instagram continues to provide users with either one or two weeks of data across regions (see \Cref{fig:instagram_cdf_regions}), following the same pattern observed across EU countries.
On the contrary, as shown in~\Cref{fig:tiktok_cdf_regions}, we find that TikTok shows more uniform behavior in the USA and Brazil (around 180 days of browsing history data is provided) compared to the erratic behavior within EU. 

\noindent

\noindent
\textbf{Important takeaways: }

\noindent
\faHandPointRight~ TikTok DDPs are the most reliable. While, Instagram failed the completeness and consistency audits for our sock-puppet accounts, YouTube failed the consistency audit.

\noindent
\faHandPointRight~ Instagram and YouTube exhibit varying degrees of missing data across snapshots. Moreover, Instagram's DDP is not complete as it does not contain viewed reels.

\noindent
\faHandPointRight~ There are disparities in the amount of data within a platform's DDPs across users within EU. Instagram and TikTok have clear variations for video browsing and search histories, while YouTube appears to maintain a more consistent data distribution by providing data for the entire history.

\section{Assessing comprehensibility of DDPs}
\label{Sec: Comprehensibility}
Despite the presence of the most reliable form of information, the utility of DDPs will be significantly undermined if the communication of such information is not comprehensible to users. 
Therefore, in this section, we investigate whether the current implementations of platforms adhere to the GDPR's comprehensibility requirements.

\noindent
\textbf{Methodology overview}: Per Article 12 of GDPR, DDPs should be in a \underline{concise}, \underline{transparent}, \underline{intelligible} and easily \underline{accessible} form, using \underline{clear and plain language}~\cite{EU2016GDPR}.
As the GDPR does not provide more succinct definitions of these requirements, 
following trends in empirical contract research 
~\cite{BenShaharContracts2017}, we conduct a large-scale user survey among participants from different European countries to understand their interpretations of some of the requirements. 
Finally, we evaluate the current implementations, i.e., the DDPs obtained from TikTok, Instagram, and YouTube, against the interpretations by participants.

\noindent
\textbf{Participant recruitment}:
To conduct our survey, we recruited 100 participants each from Germany, France, Spain and Italy (400 in total) who had a high approval rate ($\ge98\%$) on Prolific~\cite{prolific2025prolific}.
Our motivation for a general set of respondents stems from the fact that the GDPR empowers general residents in the European Union (EU) with the Right to data access, and requirements are intended to benefit these residents. 
Though, eventually, our plan is to extend the study to \emph{all} countries within the EU, we chose to start with the countries that have the highest GDP~\cite{statista_gdp}.
%
We also selected the standard gender breakdown while setting up the surveys.
In total, we recruited 228 males (57.0\%), 161 females (40.3\%), 9 self-reported as ``other'' (2.3\%), and 2 participants who preferred not to disclose their gender (0.5\%) (See \Cref{Tab: Demographics} in Appendix \ref{appendix:survey} for details).
Prior to entering our survey, we presented an online consent form for expressing explicit consent from the participants. 
Upon completion of the survey, we compensated each participant at a rate of \pounds 9 per hour, which is recommended by Prolific to be a good and ethical rate of remuneration~\cite{prolific_remuneration}.

Our survey comprised of three components: (a)~awareness of participants about the rights, (b)~their interpretation of the requirements and (c)~adherence evaluation.
We elaborate on the survey setup and important observations for the three components.  



\subsection{Awareness about the GDPR Right of Access}
\label{appendix: Awareness}

\noindent
\textbf{Survey setup}: In the first component, we first tried to educate our participants about Article 15 of the GDPR and how they can request their data from online platforms. 
We also provided a link to a Google Drive folder where they could see an example of a DDP. 
At this stage, we asked our participants three questions regarding (1) whether they were aware of such right before taking our survey, (2) whether they had exercised this right by requesting their data, and (3) if they had exercised their rights, then why.
We designed this part of the survey to understand the general awareness of the participants and make them aware of the said right. 

\noindent
\textbf{Observations}: Out of the 400 participants, nearly $72\%$ participants responded that they were aware of the Right of Access to data before participating in our survey.
At the same time, only $29.2\%$ of them answered in affirmation when it comes to exercising their rights by requesting the data on some platform. 
Although there is a massive gap between awareness and exercise of the data requests, the numbers are surprisingly high.
We also asked our participants to provide reasons for exercising their data requests in free-form tex.
Through manual annotations the responses were characterized into four groups. 
Almost half of the participants (51\%) mentioned \textit{curiosity}, knowing what information platforms collect about them, to be the primary reason for their data request.
Further, 18\% mentioned \textit{seeking specific information}, e.g., identifying a song they had listened to earlier, cross-checking specific details, or determining the amount of time spent on the platform, etc., as the primary reason for their requests.
Finally, 11\% of our participants made their requests to \textit{keep a backup}.

Interestingly, one-fifth of the participants (20\%) mentioned that they requested data from several platforms for participating in some research study. 
Note that several recent studies--for understanding social media-- use data donations from users on Prolific (\cite{yang2024coupling}, \cite{vombatkere2024tiktok}) and other crowd-sourcing platforms as their primary data source.
This observation likely explains the surprisingly high percentage of awareness and exercise statistics that we reported above.


\subsection{Interpretation of the requirements}
\label{Sec: ComprehensibilityDesiderata}


\noindent
\textbf{Survey setup}: To understand the interpretation of requirements by common end users, we showed the participants the relevant paragraph of Article 12, which mentions the different requirements of DDPs. 
We asked our participants about their interpretation or expectation of the shared personal data in the DDPs to follow any of these requirements.
To understand their interpretations of different requirements, we utilized BERTopic~\cite{grootendorst2022bertopic}.
BERTopic clustered similar interpretations together into `topics'; we then manually \changes{derived themes based on keywords and representative sentences to understand the underlying abstract nuances.}
Next, we elaborate on the observations in the context of each comprehensibility requirements.

\noindent
\textbf{Conciseness:}
The important themes that resonated across the responses spanning all countries are -- prioritize important \& relevant data, summarize important observations, and present data in a structured way that is easier to navigate. 
We observe an interesting difference in the interpretations across countries. 
Participants from Italy, France and Spain mentioned the documents should be shorter, while others mentioned avoiding an overwhelming amount of text. 
In contrast, participants from Germany only prioritize the informativeness of the content irrespective of its length. 

\noindent
\textbf{Transparency:}
Unlike the case of conciseness, there are no major differences in how participants from all countries interpret transparency. 
The interpretation of transparency can be divided into two major aspects. 
First, participants call for unambiguous, full disclosure of all the data that has been collected about them in the way they are collected. 
Second, and most importantly, some participants also expect the DDPs to mention how the data is processed and for what purpose the data is collected. 

\noindent
\underline{\textit{Conciseness vs. Transparency:}} As per participants' interpretations, conciseness and transparency requirements are at odds with each other. 
Therefore, any DDP representation may not satisfy these requirements simultaneously, i.e., to satisfy one, the other has to be (at least partly) disregarded.

\noindent
\textbf{Intelligibility:}
All the participants explicitly interpret it to be easy to understand and link it closely with the language being straightforward and without any technical or legal jargon. 
Participants from all countries also mentioned that intelligible data should have a clear explanation of how and why a certain piece of information is collected.
To this end, participants also noted some interesting expectations. 
For instance, participants from Germany mentioned that platforms should share DDPs in one's language of choice. 
Moreover, participants from Italy stated that platforms should share their DDPs in the form of an elegant visualization. 

\noindent
\textbf{Clear and plain language:}
Most of the participants interpret plain and simple language to be words that are easily understandable and unambiguous. Some participants elaborate on the nuance by stating that legal and technical terms should be avoided or at least explained in a simpler way. 
To this end, we observe that participants do not distinguish between intelligibility and language requirements.

\noindent
\textbf{Accessibility:}
Most of our participants interpret accessibility as being easily available, i.e., it is easy to find the place to request the data and to access the data.
However, there are certain interesting and valid interpretations that are worth mentioning. 
Some participants mentioned that the data needs to be available in a format so that it can be opened without any sophisticated software. 
Also, the timeliness with which the data is provided upon request is an interesting aspect mentioned by participants. 
Another important interpretation was to make sure that users with visual impairments should be able to access their data using features like screen readers.

\noindent
\textbf{Interpretations by EDPB: }
To understand whether participants' interpretations align with that of legal practitioners, we refer to the European Data Protection Board's (EDPB)~\cite{EDPB2018Transparency} guidelines which is to be followed by platforms while implementing the requirements. 
Along with agreeing on most of the above interpretations, EDPB also adds some more nuanced interpretations in its guidelines.  


While both participants and EDPB interpret concise presentation to be succinct presentation to avoid information fatigue, EDPB also asserts platforms should differentiate between privacy sensitive and non-sensitive information in DDPs. 
Similarly, in addition to the mentioned interpretation of transparency, EDPB also asks platforms to clearly mention the most important consequences and risks that each data processing category may entail. 
Much like the tension we found between participants' interpretation, even EDPB recommendations reflect such tensions between conciseness and transparency. 
For intelligibility, EDPB completely agrees with participants' interpretations by linking it to the usage of clear and plain language which could be understood by an average member of the intended audience. 
Along with understandability, for the interpretation of clear and plain language, EDPB recommends using definitive language that does not leave any room for different interpretations. 
Finally, for accessibility, EDPB's recommendations completely match that of our surveyd participants. 


\subsection{Adherence evaluation}
\label{Sec: AdherenceEval}

\begin{figure*}
    \centering
    \includegraphics[width=0.7\textwidth,height=7cm]{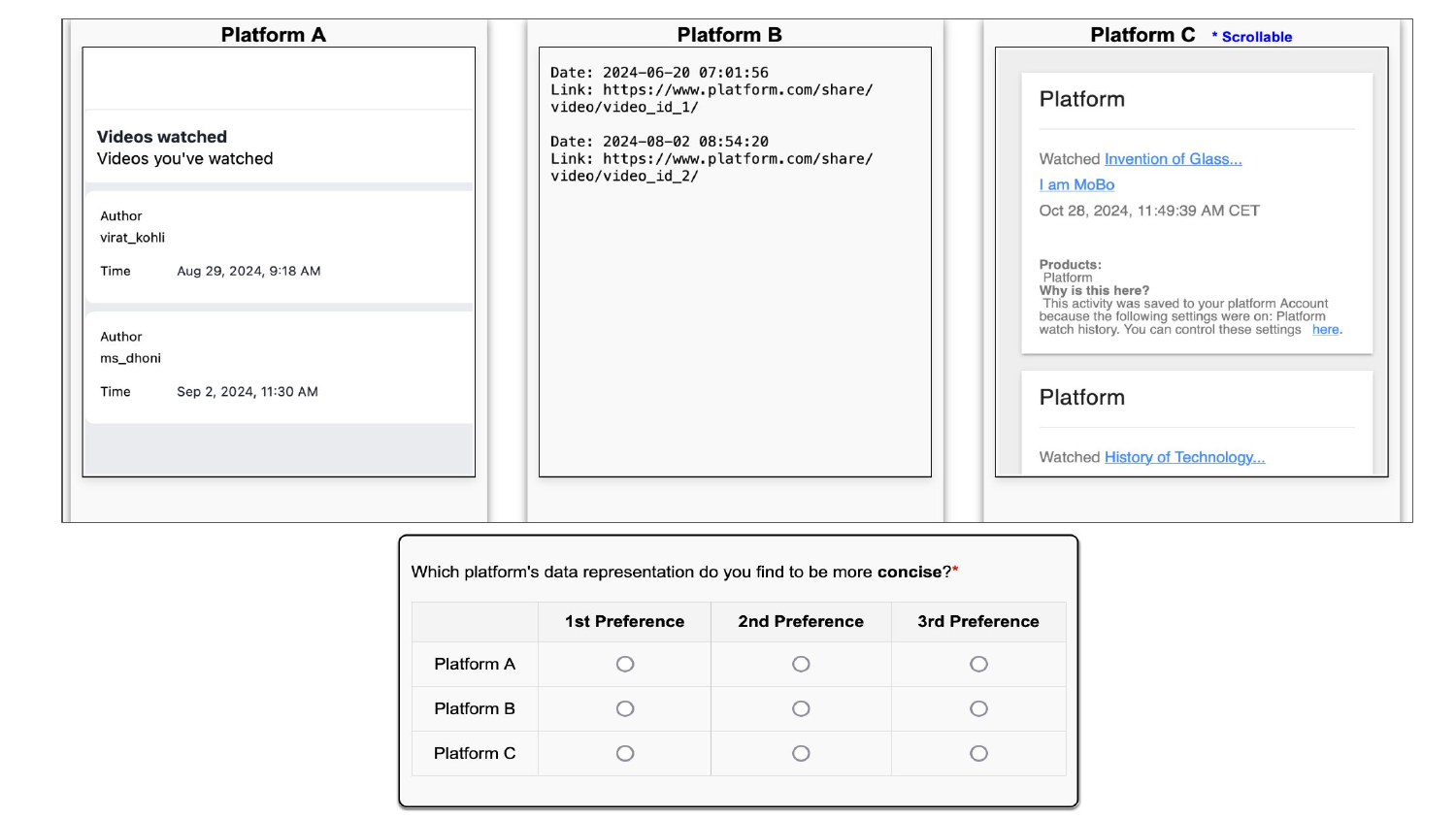}
    \caption{Figure representing how the watch history from the three platforms was displayed to users for comparison and to evaluate the four properties: conciseness, clear and plain language, intelligibility, and transparency.
    }
    \label{Fig: comparitive_study}
\end{figure*}

\begin{table*}
\small
\begin{tabular}{@{}ll@{}}
\toprule
\textbf{Category}     & \textbf{What does it mean?}                                                                                            \\ \midrule
Watch history         & List of contents watched by a user on the platform.                                                                    \\
Search history        & List of terms/users searched by a user on the platform.                                                                \\
Comments              & List of comments written by a user on others posts/videos on the platform.                                             \\
Saved                 & List of contents saved by a user for future reference on the platform.                                                 \\
Connections           & List of other users connected to a user on the platform.                                                               \\
Location              & The most recent location as recorded by the platform.                                             \\
Autofill information  & Personally Identifiable Information including name, mail id, phone number, address etc. often collected implicitly.    \\
Off-platform activity & User's activity data shared by other platforms with the platforms under consideration.                                 \\
Devices               & Details of the devices that a user has used to connect to the platform.                                                \\
Login history         & Device and network from which the user has logged into the platforms.                                                  \\
Like history          & List of contents liked by a user.                                                                                      \\
Personal information  & Personally Identifiable Information, including name, mail ID, phone number, address, etc., shared by the user. \\
User's content        & List of contents uploaded or posted by the user on the platform.                                                       \\ \bottomrule
\end{tabular}
\caption{A list of categories of data that were considered in the adherence evaluation phase of the user survey.}
\label{Tab: FieldsEval}
\end{table*}

\noindent
\textbf{Survey setup}: In the final part of the survey, we presented our participants with the information shared in the DDPs of TikTok, Instagram and YouTube in an anonymized and randomized format \changes{as shown in~\Cref{Fig: comparitive_study}}. We showed different categories of content and the whole DDP to the participants.
~\Cref{Tab: FieldsEval} 
summarizes the categories that we evaluated.\footnote{Note that YouTube does not provide like history, autofill, location, and off-platform activity. Hence, for these categories, we asked our participants to evaluate between Instagram and TikTok.}

For each category, we showed participants the content shared by the three platforms side by side and asked their preference of data representation they found to be more concise/intelligible/transparent/in clear and plain language. 
We posed this question as a multiple-choice grid \changes{(See \Cref{Fig: comparitive_study})}, so each participant had to provide an ordered preference. 
In this context, first preference means the most preferred choice of data representation, whereas the third preference means their least preferred choice. 

Note that the final part of the survey consisted of 14 (13 categories + overall DDP) $\times$4 evaluations where participants gave us an ordered preference. 
Hence, to avoid participant fatigue, we asked half of our participants to evaluate intelligibility and transparency and asked the other half to evaluate conciseness and language properties.
Hence, 200 participants answered each of the evaluation questions.
Next, we report our findings on the adherence evaluation.

\begin{figure} 
    \centering
    
    \hfill
    \begin{subfigure}[t]{0.49\columnwidth}
        \centering
        \includegraphics[width=\textwidth, height=3cm]{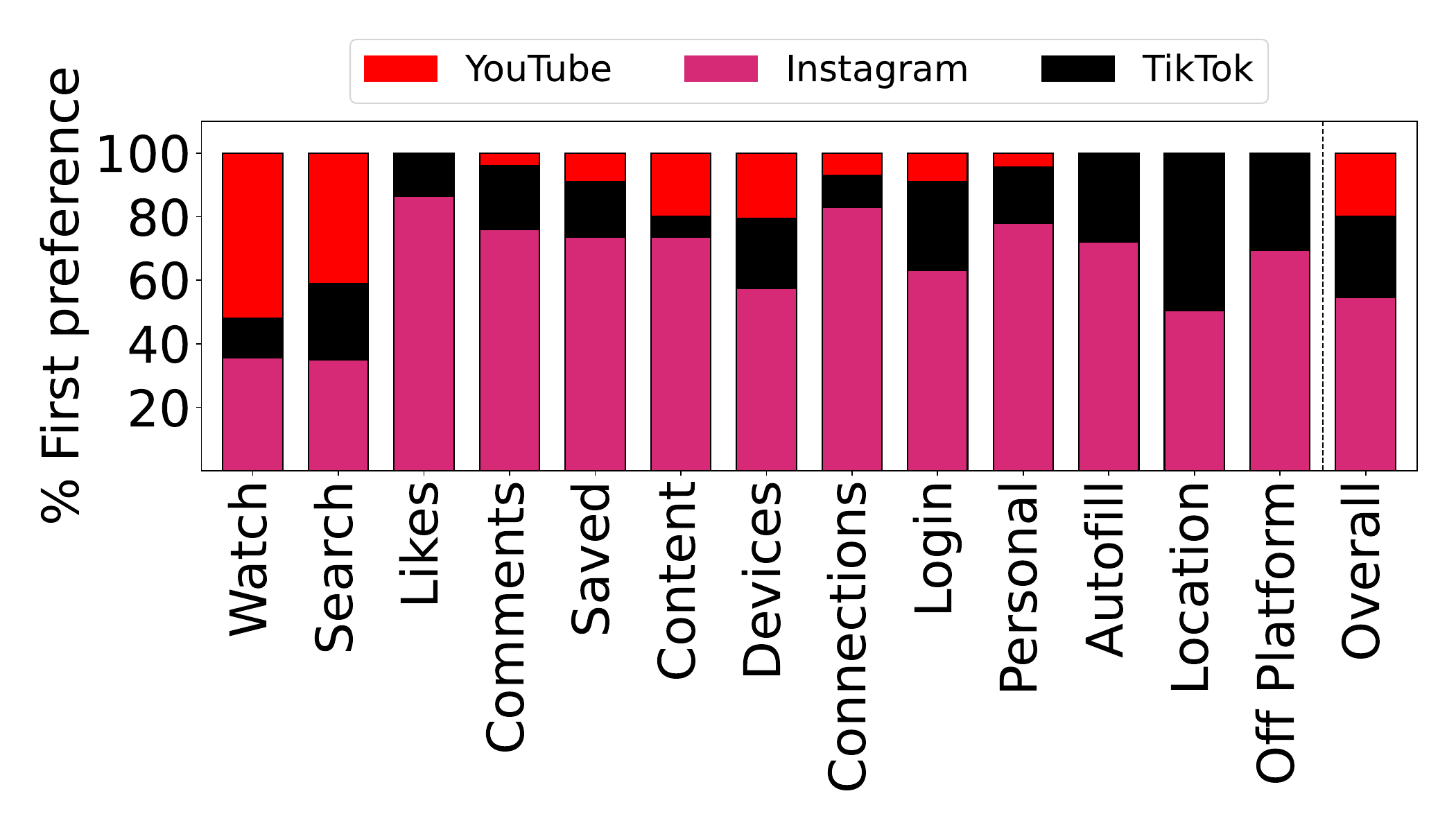}
        \caption{Clear and Plain Language}
        \label{fig:language_first_preference}
    \end{subfigure}
    \begin{subfigure}[t]{0.49\columnwidth}
        \centering
        \includegraphics[width=\textwidth, height=3cm]{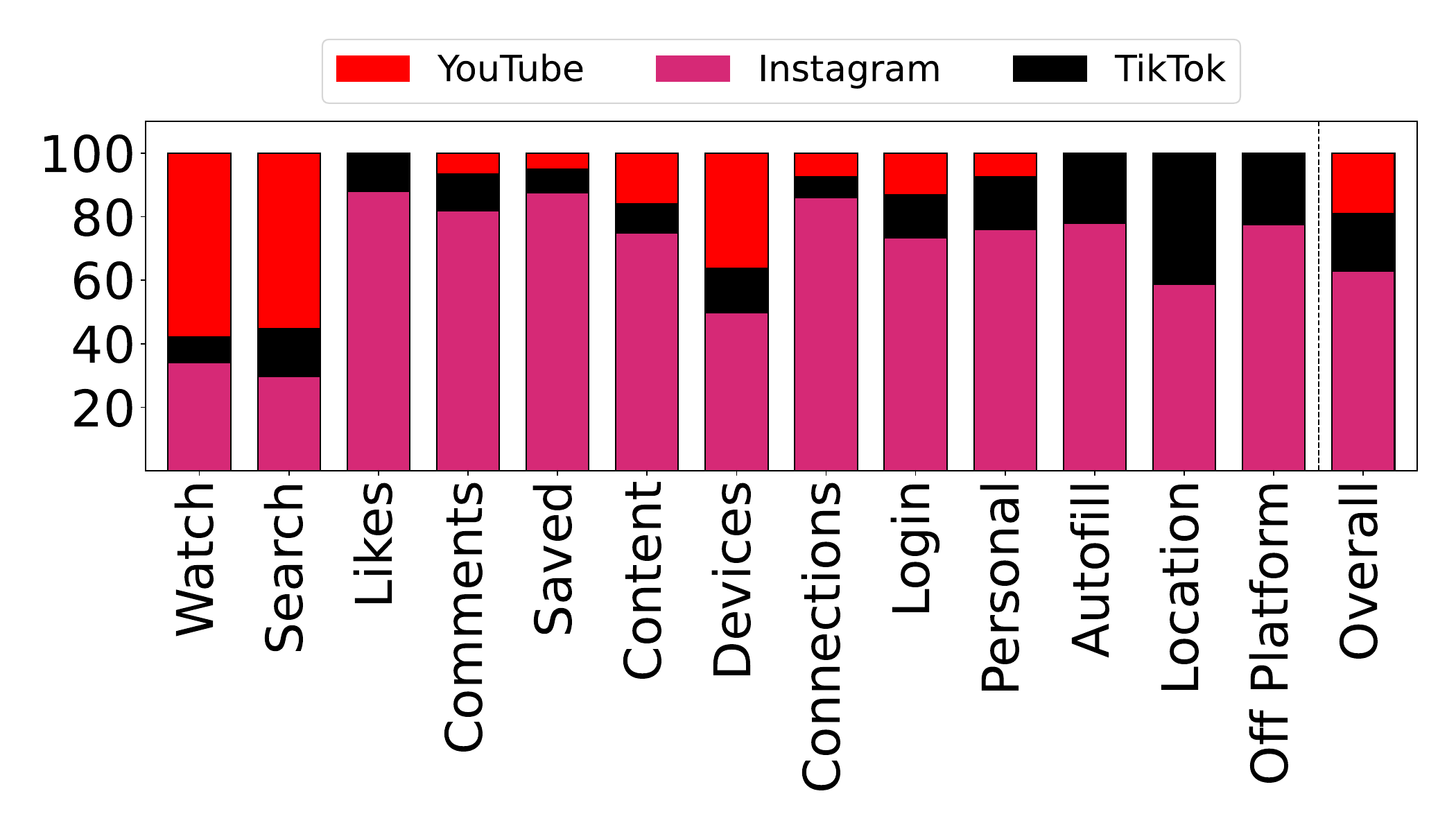}
        \caption{Intelligible}
        \label{fig:intelligible_first_preference}
    \end{subfigure}
    \begin{subfigure}[t]{0.49\columnwidth}
        \centering
        \includegraphics[width=\textwidth, height=3cm]{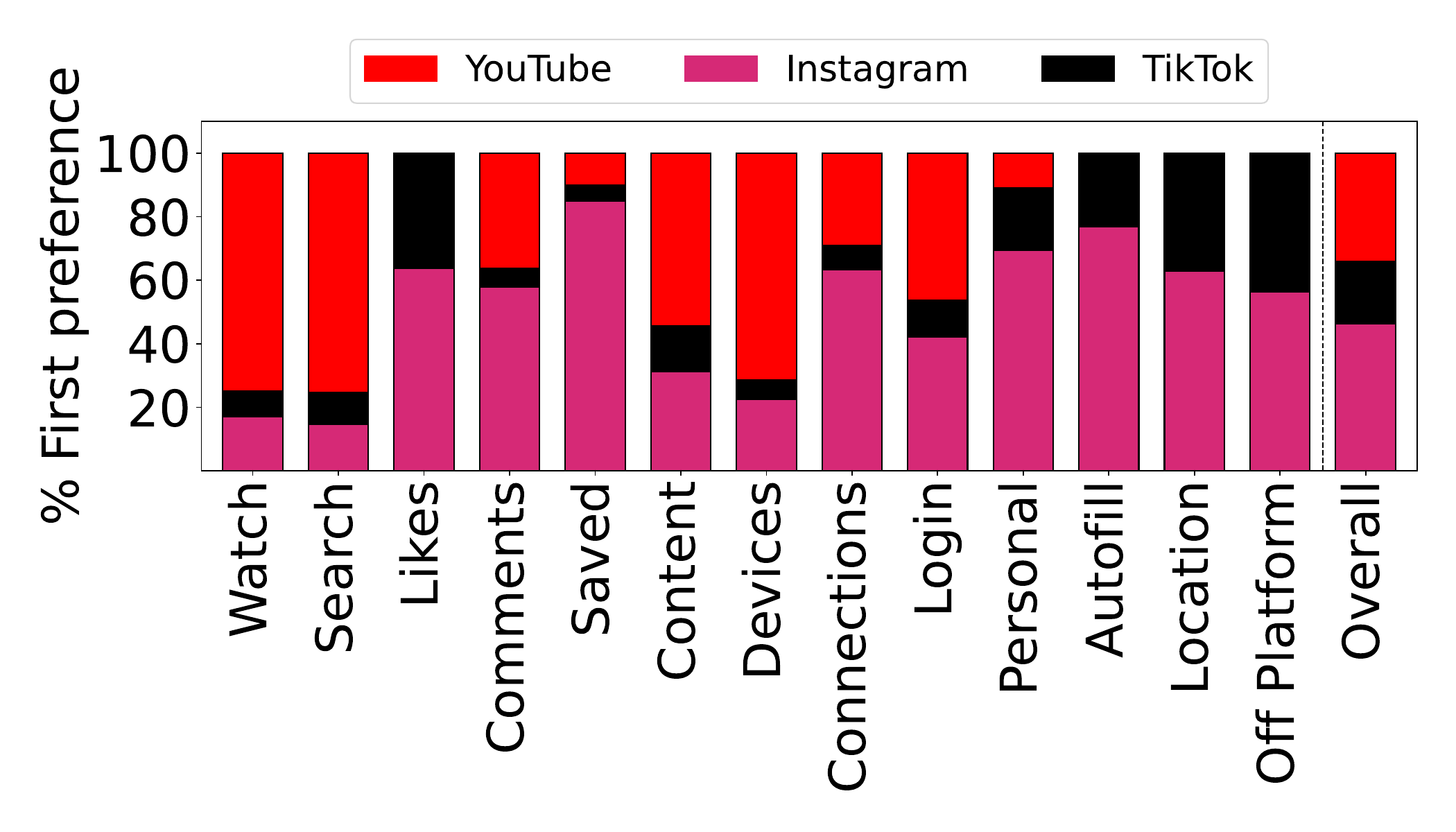}
        \caption{Transparency}
        \label{fig:tranparency_first_preference}
    \end{subfigure}
    \hfill
    \begin{subfigure}[t]{0.49\columnwidth}
        \centering
        \includegraphics[width=\textwidth, height=3cm]{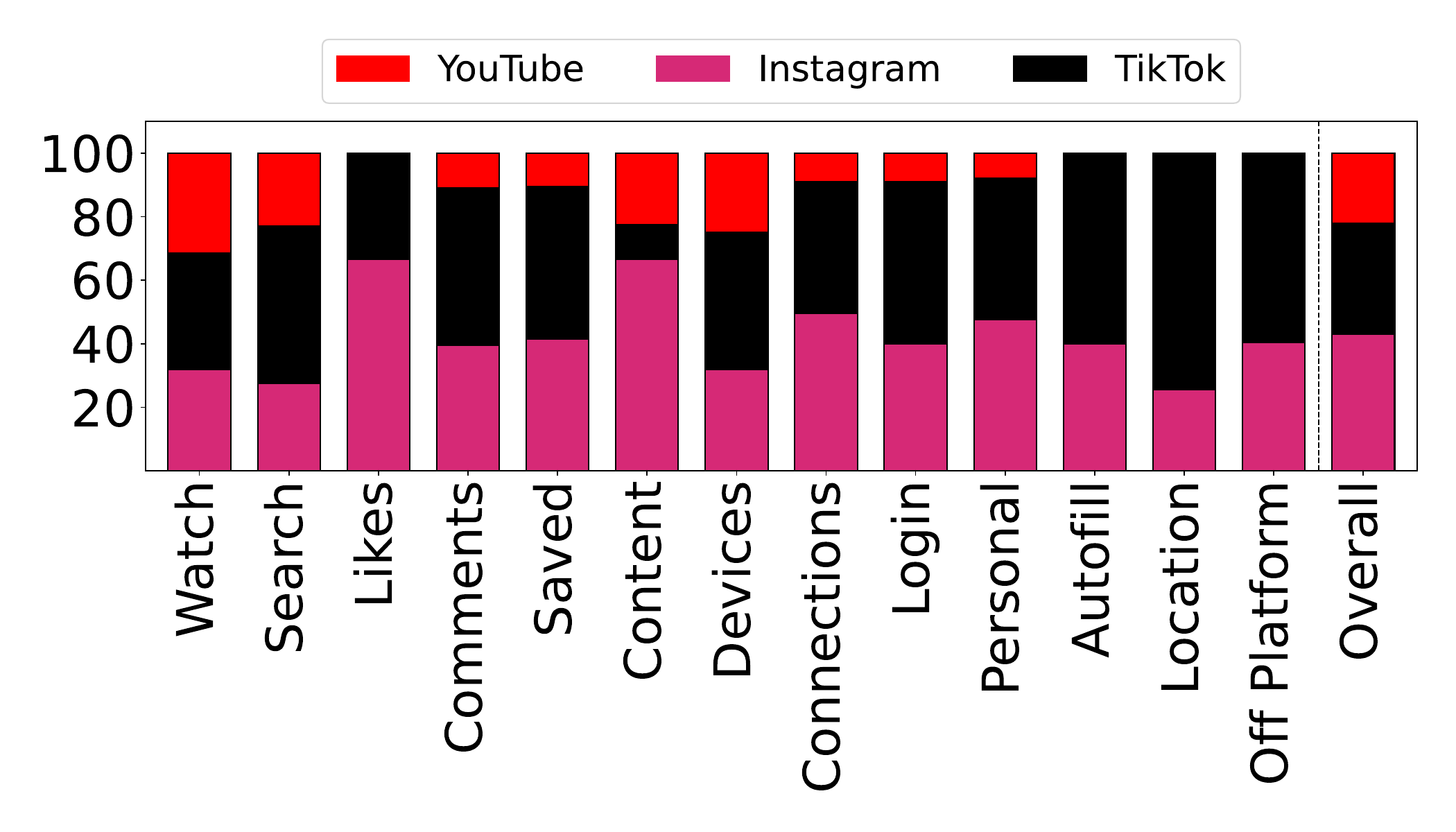}
        \caption{Conciseness}
        \label{fig:concise_first_preference}
    \end{subfigure}

    \caption{Percentage of first preferences across all categories and for the entire DDP. Instagram is the top choice for most categories across all requirements, except for conciseness. For watch and search histories, YouTube is the most preferred platform in transparency, intelligibility, and language.}

    \label{Fig: ComprFirstPref}
\end{figure}

\noindent
\textbf{Language}: \Cref{Fig: ComprFirstPref} shows the percentage of first preference votes elicited by the representation of each of the platforms for the four requirements that we surveyed.
In the evaluation of clarity and simplicity of language, Instagram won the first preference votes in eleven out of the thirteen data categories (\Cref{fig:language_first_preference}). 
In contrast, for the remaining two categories - search and watch history - participants preferred the language in which YouTube presents it.
%
Apart from the thirteen data categories, we also asked participants to evaluate the overall DDP from the three platforms. 
55\% of the participants preferred the language of Instagram's DDP. 
TikTok and YouTube DDPs are preferred by 25\% and 20\% of the participants, respectively.

\noindent
\textbf{Intelligibility}: Similar to Language, Instagram won the first preference vote for intelligibility in eleven out of the thirteen categories evaluated (\Cref{fig:intelligible_first_preference}).
In all categories, except for devices and location categories, more than 75\% of the participants preferred Instagram's data representation to be the most intelligible.
For the evaluation of the overall DDP, 63\% of the participants identified the Instagram DDP to be the most intelligible.
TikTok and YouTube were chosen as the first preference by a considerably smaller proportion of our participants, nearly 18\% each.

\noindent
\textbf{Transparency}:
YouTube won the first preference for transparency requirement in five categories, where over 70\% of the participants chose it for watch
history, search history, and login history, followed by devices and content with 55\% and 46\%, respectively
For the remaining eight out of the thirteen categories, Instagram won the first preference of the participants in a large percentages.
For the overall DDP, Instagram is preferred more compared to the other two platforms, with nearly 46\% of the participants choosing it to be the most transparent.
YouTube and Tiktok are preferred by 35\% and 19\% participants respectively.


\noindent
\textbf{Conciseness}: Based on~\Cref{fig:concise_first_preference}, we found that in nine out of thirteen categories of data that we surveyed, respondents preferred TikTok's data representation as the most concise, whereas Instagram's data representation is found to be the most preferred for the remaining four categories. While TikTok's representation is preferred in the location, autofill, and off-platform categories ($\ge$60\% first preference votes), Instagram's representation is preferred more for likes history and contents ($\ge$65\% first preference votes).
While evaluating the overall DDP, 43\% of the participants found Instagram's DDP to be more concise, followed by Tiktok and YouTube with 35\% and 22\%, respectively.

\noindent
\textbf{Accessibility}: 
We excluded accessibility from the survey evaluation because it pertains to getting access to request and download DDPs more than to their representations.
Since it is the most objective requirement for evaluation, we note down the observed differences across platforms below.

In terms of \textit{ease of request}, from the content page, one needs to click \textbf{6}, \textbf{6}, and \textbf{10} times on TikTok, YouTube, and Instagram, respectively, to be able to request the data in the human-readable format. 
Notice that, while on Instagram, a user needs to click more, the platform provides many options to the end-user, such as the time duration.
Based on our anecdotal observations, the \textit{turnaround time} for TikTok upon request was instantaneous during this study.
For Instagram and YouTube, it often takes 10 to 15 minutes for the DDP to become available.
Thus, all the turnaround times are within the GDPR-prescribed maximum duration of \textit{one} month.

One can download one's data on TikTok and Instagram from both their applications and web versions. 
On the other hand, YouTube shares the DDP by sending an e-mail with a URL to it.
All these platforms provide DDPs in a compressed format, which makes it harder for users to open them on a smartphone.
At the same time, a user needs to have a browser or text reader to \textit{open} the HTML (YouTube and Instagram) and TXT (TikTok) DDPs.

In summary, while the data request and download are relatively accessible in the current setting, reading the data is still complicated for the average end user.
At this point, we would also like to note that we are not aware of any special provisions that platforms may have for people with visual impairments to access the data.

\noindent
\textbf{Variations across countries}: For the survey based evaluation, we do not observe any qualitative variation across the four countries for three requirements -- language, intelligibility, and transparency. 
However, for conciseness, participants from Germany predominantly preferred Instagram, with over 50\% of the participants selecting it as their first choice across most categories. 
In contrast, participants from other countries selected TikTok as having the most concise representation for most of the categories.

Upon a manual inspection of the interpretation of conciseness, we found that, except participants from Germany, most other users interpret conciseness as: (1) minimizing document length, (2) avoiding information overload etc. 
This difference in interpretation is reflected in how our participants evaluated the data representations. \\

\noindent
\textbf{Important takeaways: }

\noindent
\faHandPointRight~We found participants' interpretation of the requirements agree with that of European Data Protection Board.

\noindent
\faHandPointRight~However, conciseness and transparency requirements of the GDPR are inherently at odds with each other.

\noindent
\faHandPointRight~People assess Instagram's current data representation to be most comprehensible followed by that of YouTube. 

\section{Improving comprehensibility of DDPs}
\label{Sec: Recommendation}

While the previous section uncover a number of shortcomings in comprehensibility of the DDPs, in this section, we present some recommendations to improve their comprehensibility. 
To this end, our proposed recommendations are two pronged (a)~improving data representation within DDPs, (b)~improving the presentation of DDPs. 

\subsection{Improving data representations within DDPs}
\label{Sec: RecommendationRepresentation}

\begin{figure}
    \centering
    \includegraphics[width=0.90\columnwidth, height=3.75cm]{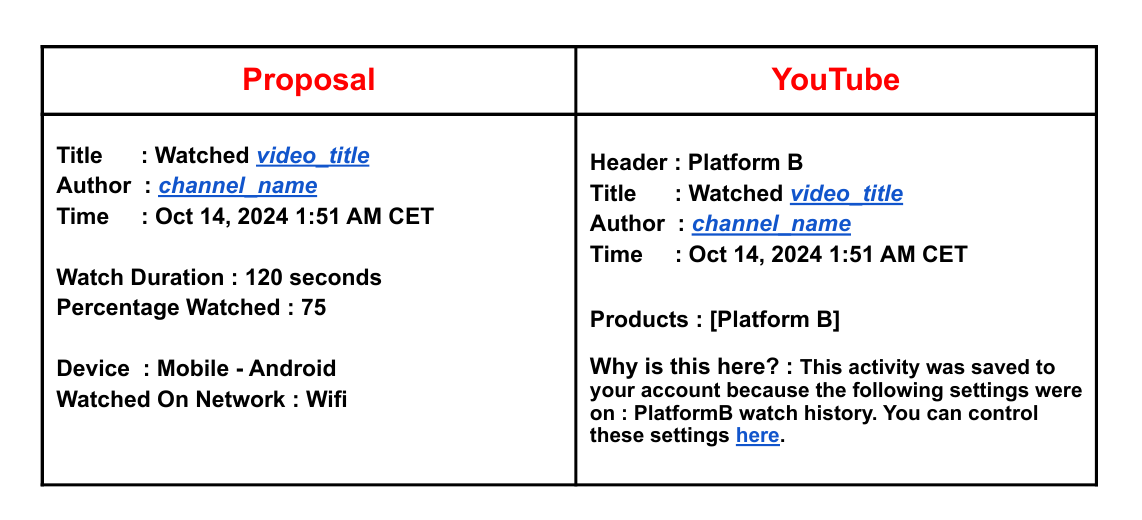}
    \caption{ Comparison of `watch history' in our proposal and that of YouTube's current DDP implementation. 
    }
    \label{fig:recommendation_representation}
\end{figure}

\begin{figure}[t] 
    \centering
     \begin{subfigure}[t]{0.49\columnwidth}
        \centering
        \includegraphics[width=\textwidth, height=3.5cm]{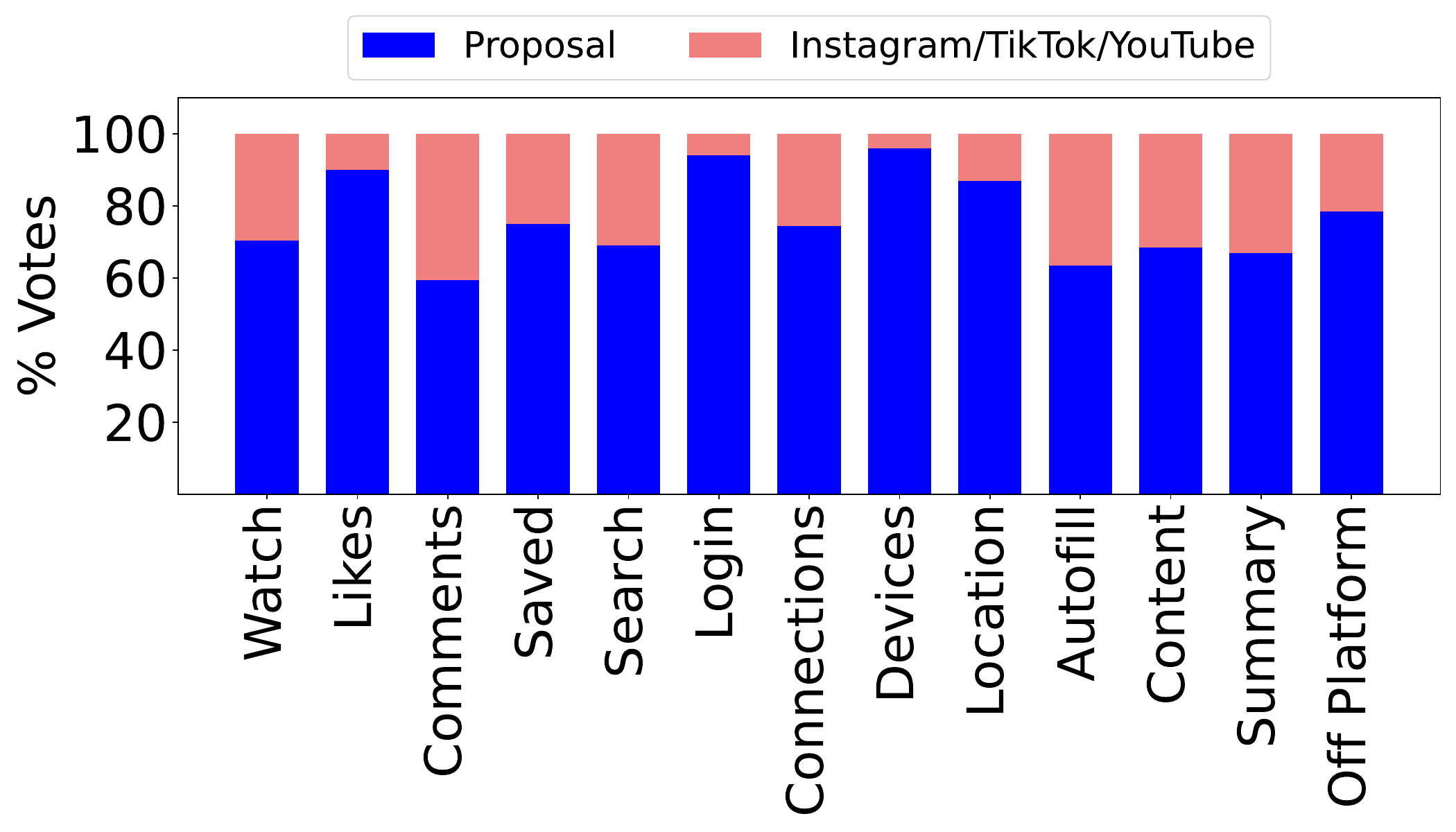}
        \caption{Clear and Plain Language}
        \label{fig:language_first_preference_our_prop}
    \end{subfigure}
    \begin{subfigure}[t]{0.49\columnwidth}
        \centering
        \includegraphics[width=\textwidth, height=3.5cm]{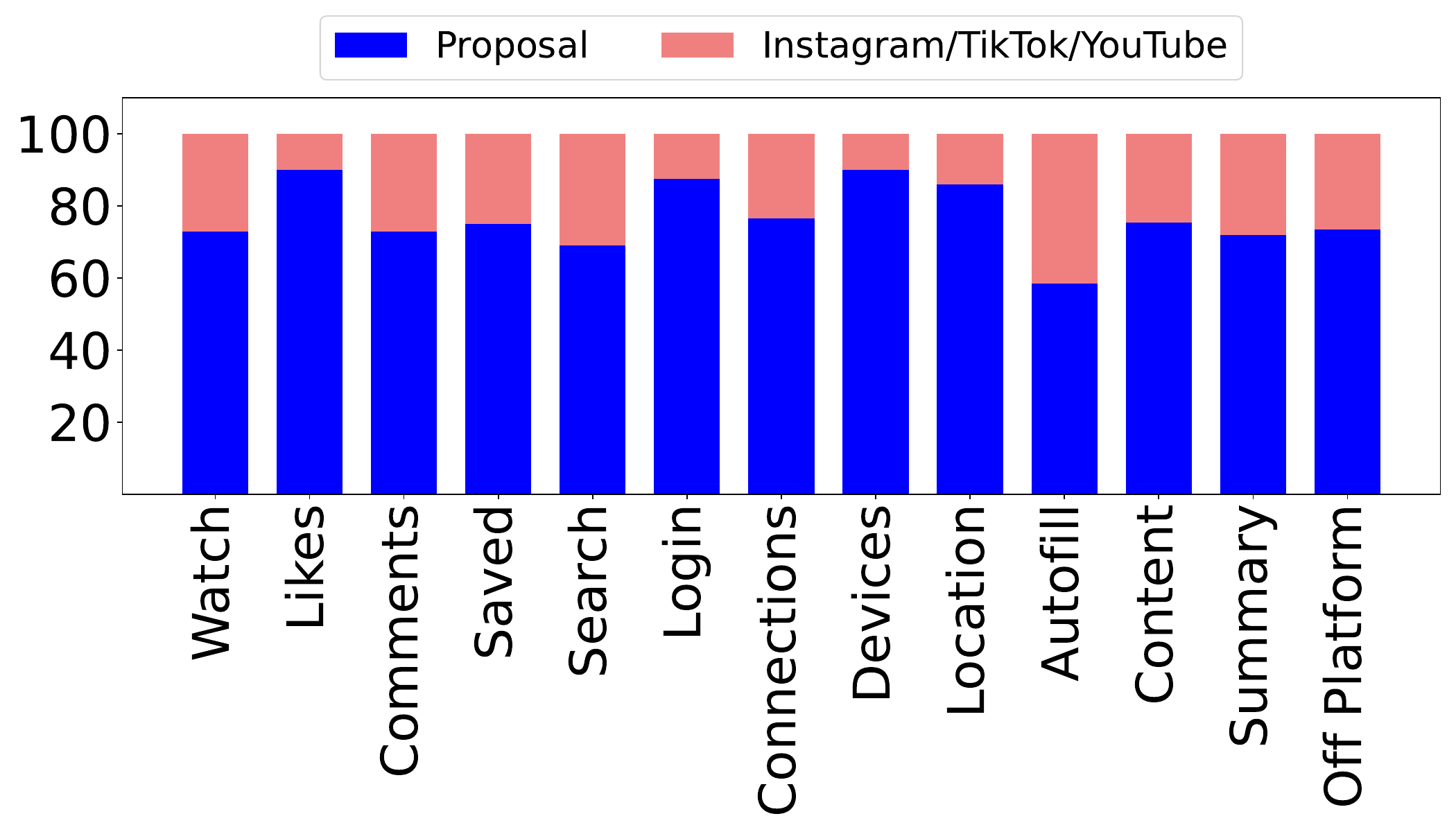}
        \caption{Intelligible}
        \label{fig:intelligible_first_preference_our_prop}
    \end{subfigure}
    \hfill
    \begin{subfigure}[t]{0.49\columnwidth}
        \centering
        \includegraphics[width=\textwidth, height=3.5cm]{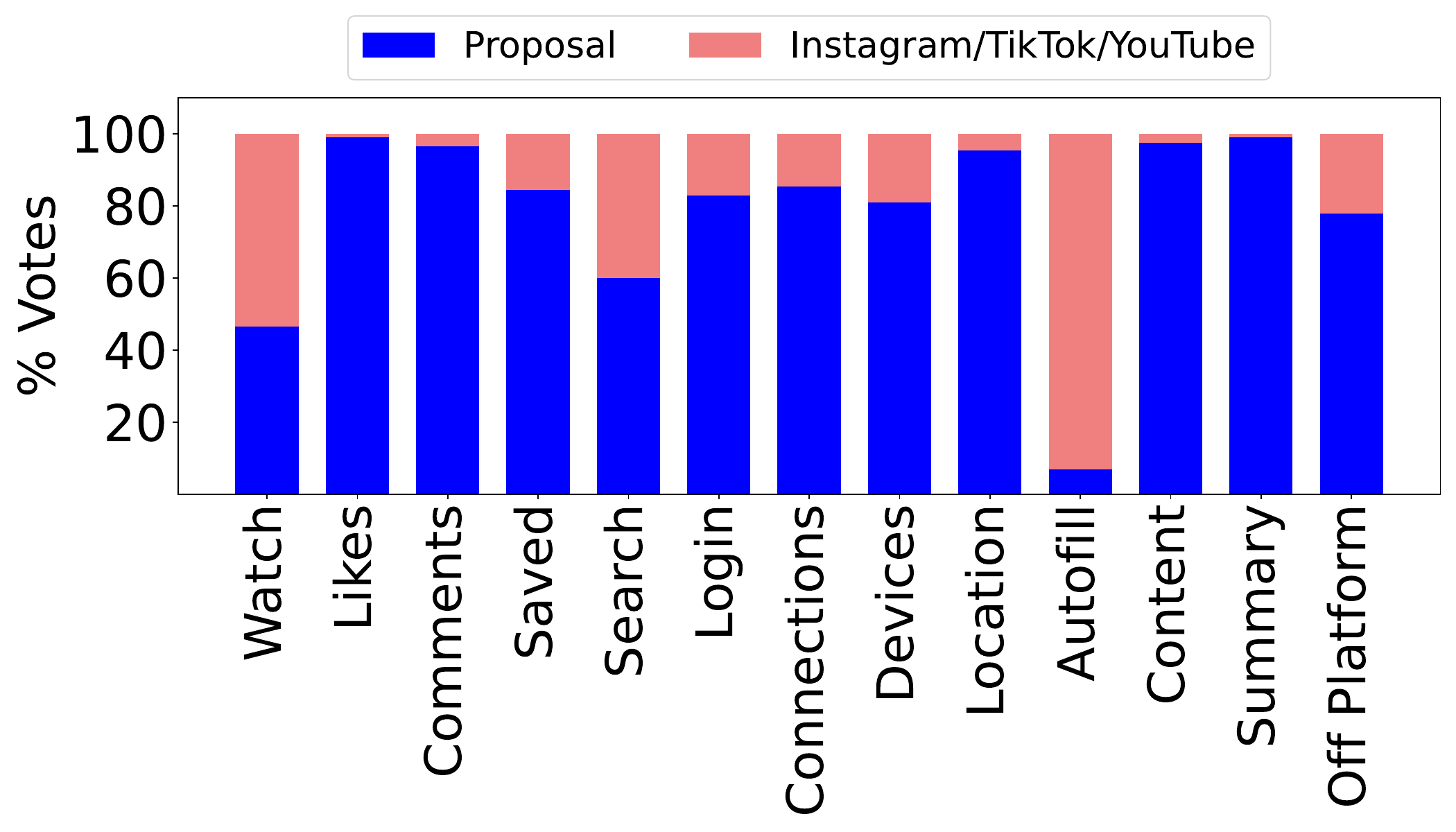}
        \caption{Transparency}
        \label{fig:tranparency_first_preference_our_prop}
    \end{subfigure}
    \hfill
    \begin{subfigure}[t]{0.49\columnwidth}
        \centering
        \includegraphics[width=\textwidth, height=3.5cm]{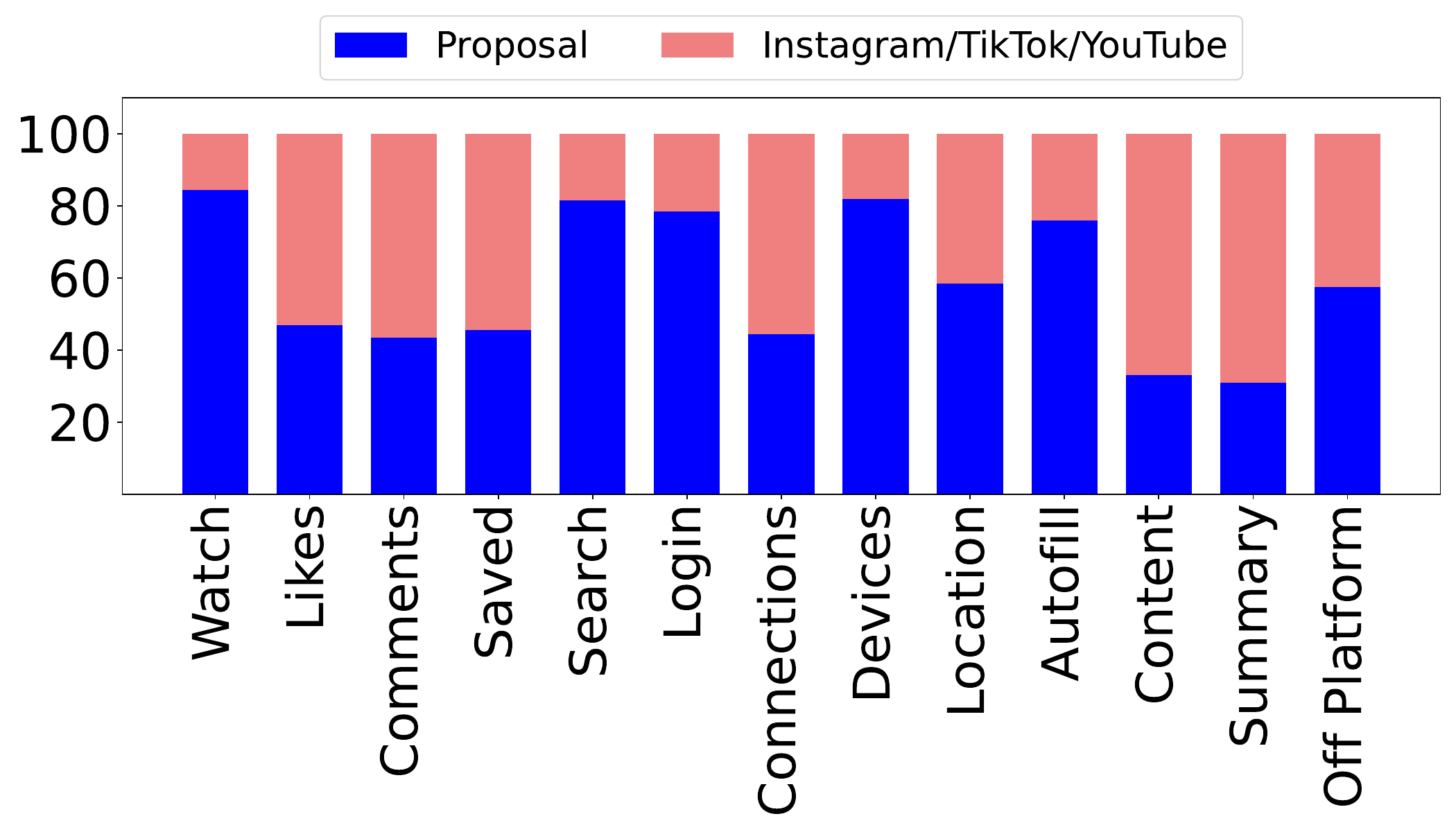}
        \caption{Conciseness}
        \label{fig:concise_first_preference_our_prop}
    \end{subfigure}
    
    \caption{ Percentage of votes for the proposal and the earlier winning platform across different categories for each requirement. Our proposal emerged as the top choice in all fields for the \textit{language} and \textit{intelligible} requirements.} 
    \label{Fig: RecoEval}
\end{figure}

To improve the data representation within DDP, we deploy an AI assisted human-in-the-loop approach. 
In this approach, we leverage the different interpretations obtained by survey participants and EDPB guidelines and use a large language model (LLM) to come up with an initial data representation. 
Then, we modify these initial representations to propose the final data representation. 

\subsubsection{Methodology}
\label{Sec: RecMethodology}
To come up with a recommendation for data representation, 
we take the following steps.

\noindent
\textbf{Step 1: Summarize the interpretations}: For each of the GDPR requirements, we collect the interpretations from our participants and EDPB guidelines~\cite{EDPB2018Transparency}. 
Digesting such a diverse set of requirements and their expectations, and ironing out data representations that adhere to such requirements and live up to the expectations of different stakeholders is cognitively challenging even for experts.
Hence, we feed this entire set of interpretations to an LLM -- Gemini 2.0 Flash \cite{google2024gemini}-- using the Google AI Studio platform. 
We ask the LLM to generate a set of representative sentences that encapsulate the most prevalent interpretations by summarizing the above responses. For reproducibility purposes, we make the prompts we used and the responses available in \changes{the GitHub repository.}
Note that in this approach, we include the interpretations of both data subjects and data protection authorities (through the EDPB interpretations).

\noindent
\textbf{Step 2: Generate recommended data field for each data category}: In the second step, we pass the representative sentences obtained in Step 1 for each of the requirements to the LLM 
to obtain the specific data fields for each data category that we should provide to the end user. 
Subsequently, the LLM provides us with a set of recommendations for data fields of information in the queried category in the form of a JSON object \changes{(available in the GitHub repository).}

\noindent
\textbf{Step 3: Manual inspection by researchers}: In the final stage, three researchers (coauthors of this paper) 
go over these recommendations to assess their quality. 
The primary goal of their assessment is to remove any instance of hallucination in the LLM responses; secondly, they wanted to ensure not to recommend some data fields that will enable platforms to collect any additional personal data. 
Next, given their expertise in understanding data donations in other platforms like Netflix, Prime video etc., we ask them to suggest any additions that were deemed relevant or removal of certain fields when they contained excessive detail in the recommended fields. 
Note that while Step 1 introduces the end users' and data protection authorities' interpretations, this step also instills the interpretations of tech experts on other platforms and researchers who want to study the dynamics of platforms using these DDPs. 
Overall these interventions affected changes in 20 out of 85 data fields across all the data categories. 
By taking these steps, we ultimately come up with the final recommendation of improved representation for different data categories.

\subsubsection{Evaluation of the proposed representation}\label{Sec: EvalProposal}
To evaluate our proposed representation, we conducted another survey among 200 participants (who have not participated in the survey in \Cref{Sec: Comprehensibility}) from Germany, France, Spain, and Italy. We follow the participant recruitment strategy as elaborated in~\Cref{Sec: Comprehensibility}. 
\Cref{Tab: Demographics} in Appendix~\ref{appendix:survey} shows the user demographics.
During the survey, we showed the participants two representations for the evaluated categories: (a)~proposed representation: which we generate from the steps mentioned in~\Cref{Sec: RecMethodology}, (b)~winner representation: for each data category, the representation among Instagram/TikTok/YouTube which won the first preference along more requirements (\changes{example, as shown in \Cref{fig:recommendation_representation}}). {
We randomly presented each pair of anonymized representations to the participants and
asked them to select the representations that they find to satisfy the requirements more. 

\noindent
\textbf{Observations:}
We compare the representations of the data category fields of our proposal and that of the best among the existing short-format video platforms for 13 different categories across four GDPR requirements.\footnote{
We are not considering personal information, as there were no suggestions for adding or removing any fields. Instead, we compare activity summary, which is provided only by TikTok.}
~\Cref{Fig: RecoEval} shows the percentage of votes for the two representations obtained in the 52 evaluations (13 data categories $\times$ 4 requirements), that we surveyed.
We observe that the participants preferred our proposed representation in as many as 44 categories.
However, their preferences vary for different requirements.

\noindent
\textbf{Language and intelligibility}: For clarity and simplicity of language and intelligibility of the representation, our proposal gets upwards of 67\% of the votes in most of the categories 
, suggesting our proposed representation is both intelligible and in clear and plain language.

\noindent
\textbf{Transparency}: Our proposal is voted to be more transparent than the best of the existing representations in 11 out of the 13 categories. 
Apart from search history, in all the other winning 10 categories, our representation gets upwards of 75\% of the votes.
However, in the autofill information category, Instagram's data representation gets nearly 93\% of the votes from our survey participants. 
While Instagram's representation contain many sensitive information, our recommendation contained a subset of them needing less privacy-intrusive data gathering.

\noindent
\textbf{Conciseness}:
Notice that in~\Cref{Sec: ComprehensibilityDesiderata}, we mentioned conciseness and transparency 
are mutually competing requirements.
We observe the same conflict in the results of this survey.
Our proposal was preferred to be the most concise representation in only 7 out of the 13 categories.
In watch history, search history, login history, and device details -- our proposed representation was found to be more concise by upwards of 75\% of the participants. 
Also, similarly to the survey presented in~\Cref{Sec: AdherenceEval}, participants from Germany have different preferences compared to those from other countries for conciseness. 
Except for autofill information, over 66\% of German participants preferred our proposed representation in all other categories. 

The primary objective of this section has been to convey how one can come up with more effective data representations by considering the interpretations of different stakeholders. 
Our observation potentially indicates that currently platforms are probably operating in silos without much deliberation and sincere effort into their implementation. 
At the same time, because of the mutually conflicting nature of the requirements, it is impossible to satisfy all the requirements with any single data representation.

\subsection{Improving the presentation of DDPs}
\label{Sec: RecommendataionComprehensibility}
While the recommended representation improves the content within DDP, due to the mutually conflicting nature of the requirements (e.g., conciseness vs. transparency), it is impossible to satisfy all the requirements with any single data representation. 
Therefore, a second layer of improved presentation, in the form of a user-centric dashboard, is needed to reconcile such trade-offs. 
Along with improving along other requirements, this approach provides users the \textit{autonomy} to choose their desired level of transparency. 


\subsubsection{Methodology}
We take the following steps to improve the comprehensibility of DDPs to end users.


\begin{figure*}
    \centering
    \includegraphics[width=0.75\textwidth, height=5.5cm]{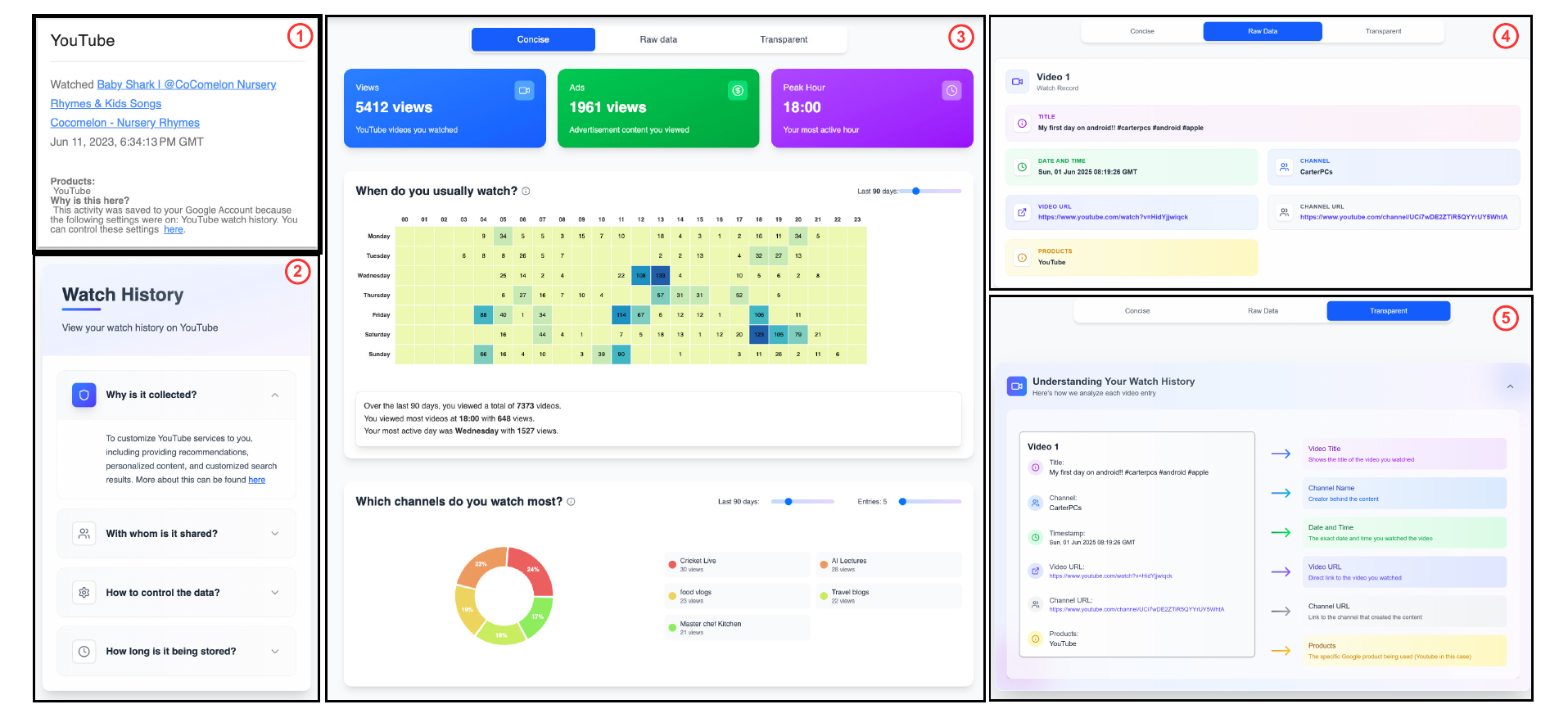}
    \caption{ \textcircled{1} shows how YouTube currently shares watch history. \textcircled{2} illustrates how our approach addresses some of the transparency expectations of our survey participants. \textcircled{3}, \textcircled{4}, and \textcircled{5} depict our proposed concise, raw data, and transparent views respectively. Refer to  \Cref{fig:recommendation_browse_tiktok}, \ref{fig:recommendation_browse_instagram} in Appendix \ref{appendix:improve_comprehensibility} for TikTok and Instagram respectively. Note that this is partial view. The full view of YouTube Dashboard can be accessed in the GitHub repository.
    }
    \label{fig:recommendation_browse}
\end{figure*}

\noindent {\textbf{Step 1: Understanding user expectations from DDPs.} 
To understand user expectations from DDPs, we conducted a structured survey focused on the data categories analyzed in \Cref{Sec: Comprehensibility}. 
A total of 120 participants were recruited from four EU countries -- Germany, France, Italy, and Spain -- with 40 participants assigned to each platform. 
Demographic details of the participants are presented in \Cref{Tab: Demographics} in Appendix \ref{appendix:survey}. 

\changes{In total, there were 1,569 responses regarding expectations across all categories and the three platforms. Using BERTopic, we clustered these responses, which revealed a broad spectrum of interests. For instance, for YouTube browsing history (see \Cref{fig:recommendation_browse} \textcircled{1}), questions ranged from high-level insights -- such as identifying the most frequently viewed channels or determining periods of peak activity -- to low-level clarifications, including interpretations of specific data fields (e.g., understanding what ``product'' refers to in a given context). Further, several recurring concerns emerged across participants, including questions about the purpose of data collection, data access, retention duration, and control over data collection. Low-level clarifications accounted for 8.85\% of the responses, while transparency-related questions (purpose, access, retention, control) contributed 17.5\%. The remaining responses were high-level insights which are summarized in \Cref{tab:user_expectations}.
}

\begin{table*}[htbp]
\centering
\renewcommand{\arraystretch}{2}
\begin{tabular}{@{}llr@{}}
\hline
\textbf{Category} & \textbf{Sample expectations} & \textbf{Percentage} \\ \hline
Watch history & 
\parbox{12cm}{Which videos I watched the most?\\
How long did I watch the video?} & 17.8 \\ \hline

Search history & 
\parbox{12cm}{How many searches do I do per day or per session?\\
What did I search?} & 11.7\\ \hline

Comments & 
\parbox{12cm}{How often do I comment?\\
On what post did I comment?} & 7.3 \\ \hline

Saved & 
\parbox{12cm}{How many saved posts do I have?\\
Do I revisit my saved posts?} & 3.6 \\ \hline

Connections & 
\parbox{12cm}{I would like to know when someone started following me.\\
I'd like to know when I subscribed to this channel.} & 8.6 \\ \hline

Devices & 
\parbox{12cm}{What devices did I use?\\
From what device and the \% of use per device?} & 4.8 \\ \hline

Login history & 
\parbox{12cm}{What times did I most log in?\\
How long did the session last for that login?} & 6.3 \\ \hline

Like history & 
\parbox{12cm}{How many likes I did?\\
Whose content do I like more?} & 5.0 \\ \hline

\end{tabular}
\caption{User expectations from different data categories in the DDPs.}
\label{tab:user_expectations}
\end{table*}

\noindent \textbf{Step 2: Designing a user-centric dashboard.} 
Based on the insights gathered from the survey, we developed an interactive and user-centric dashboard that empowers individuals to explore their data in a manner aligned with their preferences. 
The dashboard offers three key modes of interaction: (1)~a \textit{concise view}:  which presents high-level insights in a simplified and digestible format, (2)~a \textit{raw data view}: which presents the data exactly as provided by the platform, but with \textit{clearer and simpler language} for better understanding; and (3)~a \textit{transparency view}: which includes field-level explanations that clarify the \textit{intelligibility} and utility of each attribute. 
Also, we answer the concerns such as the purpose of data collection, data retention duration, and who has access to the data by collecting relevant information from platforms' privacy pages. 
\Cref{fig:recommendation_browse} shows the different types of views we proposed for YouTube watch history.

\noindent \textbf{Step 3: Delivering a browser extension tool. }
We package it in the form of a browser extension: \textit{Know Your Data}. The code is made publicly available.
In addition, the extension also automates the process of requesting and downloading data from platforms thereby improving \textit{accessibility}. \changes{The browser extension was implemented using React.js \cite{reactjs} for the frontend interface, Vite \cite{Vite} as the build tool, and Recharts \cite{recharts} to enable interactive visualizations. Performance profiling using Chrome’s task manager indicated CPU utilization below 2\% when idle, averaging approximately 25\% during typical processing, with transient peaks up to 110\% during word cloud generation.} This browser extension-based solution has key advantages in the following aspects:   
(1)~\textit{Privacy}: Significantly more privacy-preserving as sensitive data never leaves the user's control or device, 
(2)~\textit{Security}: Reduces the risk of data breaches on our end, as we never handle or store the user's DDP on our servers, 
(3)~\textit{Trust}: Users are generally more comfortable with extensions that process data locally,
(4)~\textit{GDPR compliance}: Much simpler from a GDPR perspective because we are not acting as a `data processor' in the sense of receiving and storing data on our infrastructure.

\noindent
\changes{\textbf{Generalizability}: The methods we proposed for improving the comprehensibility of the DDPs are generalizable to other platforms. To improve the representation, we first summarized the user interpretations which are platform independent and generalize as-is across platforms. We then identified platform specific data categories and incorporated them into the LLM prompt. These categories can be easily extracted by inspecting the DDP of each platform, but they generalize to DDPs of all users of a given platform. Finally, researchers who have expertize in understanding DPPs manually reviewed the LLM-generated responses. The corrections required were minimal, as the LLM was already able to produce sufficiently accurate DDP representations. 
}

\changes{For improving the presentation of DDPs, we surveyed users to know their expectations from the DDPs. These expectations can be generalized to other platforms, particularly transparency-related questions, whereas the high-level insights and clarifications depend on the type and granularity of data each platform collects.
}

\noindent
\textbf{Important takeaways: }


\noindent
\faHandPointRight~Our recommended data representation, that considers interpretation of different stakeholders, was favored by surveyed participants in 44 out of the 52 evaluations conducted across various pairs of data categories and requirements.

\noindent
\faHandPointRight~Simply improving data representation does not improve adherence to Article 12 requirements of the GDPR since some of the requirements are at odds with each other.

\noindent
\faHandPointRight~Understanding user expectations and giving them the autonomy may be an effective way to reconcile the trade-offs between the requirements. 
Our proposed \textit{Know Your Data} extension is one such operationalization.

\section{Ethical considerations}
\label{sec:ethics}
\vspace*{-9pt}
We acquired approval from the Ethical Review Board (ERB) of our university to conduct this research. 
We obtained explicit consent of each participants before they participate in our surveys. 
Importantly, we refrained from collecting any PII from our participants during the surveys or from their DDPs.
Although DDPs may contain PII, they are usually stored within specific keys or files.
To refrain from collecting them, we inspected the DDPs of our own accounts on the three platforms and collected a predefined list of keys and files that contain any PII.
Next, we developed scripts that remove these specific keys and files inside a DDP on the user's end, before collecting the modified DDP to our server.
Furthermore, based on our ERB, we would not share the donated DDPs with any third parties, and we will delete them within 3 years of completion of our study.


Finally, in~\Cref{Sec: Reliability}, we used bot accounts that interacted with publicly available content on the platforms.
We further took steps to minimize the consequences of running these bot accounts.
Specifically, we distributed the overhead of our experiment over multiple short browsing sessions.
Furthermore, even though the Terms of Service of the studied platforms prohibit the usage of automated bots, we believe that the benefits of our study outweigh any potential associated unintended consequences.
\section{Concluding discussion}
\label{Sec: Discussion}


\noindent
\textbf{Summary: }
To the best of our knowledge, this is the first systematic audit of the implementation of GDPR's \textit{Right of Access} from the lens of reliability and comprehensibility. 
By auditing the implementations across platforms offering similar services, we uncover a number of platform-specific and systemic instances of non-compliance. 
While periodic audits can ensure reliability compliance, to improve the comprehensibility compliance we also recommend a two-layered approach: (a)~an AI assisted human-in-the-loop approach to improve DDP representation, (b)~a \textit{Know Your Data} extension to improve DDP presentation. 
Our recommendations not only aim at striking a better balance between GDPR's conflicting comprehensibility requirements, but also provide more \textit{autonomy} to users to choose their desired level of transparency in data visualization. 

\noindent
\textbf{Limitations of the current work}: 
Our study, while comprehensive, has several limitations.
Our survey participants, recruited via Prolific, may not fully represent the broader spectrum of digital literacy or demographic diversity of the European Union. 
However, note that our work is comparative by nature. Hence, we do not expect the choice of participants will impact the findings presented in the paper. 
Further, our study focuses on data shared by platforms upon request and \textit{not} the data that is collected. 
Consequently, our findings on transparency of data provided are based solely on what platforms choose to share in DDPs, which might not reflect the entirety of data they process.
Including this information may further enrich the quality of the work.

\noindent
\textbf{Recommendations for stakeholders: }
First, platforms should improve the completeness, correctness and consistency in data granularity, retention periods, and included metadata across related data categories for a more effective and user-centric implementation. 
Furthermore, platforms should include the purpose of processing, retention periods, and third-party recipients of collected data. 
Such inclusion can not only improve the legal compliance but also match the user expectations from the DDPs (as observed in our survey for designing the dashboard).
Finally, as shown in \Cref{Sec: Recommendation}, rather than deciding on behalf of users, platforms should design solutions that empower users with more autonomy to make informed decisions. 

While our findings of non-compliance provide regulators with starting points to investigate the platforms, our observation on conflicting requirements should nudge regulators to think through the requirements they mandate for implementations. 
In fact, such rights will only be useful if data protection authorities provide more detailed technical specifications and/or standardization for the implementation and enforcement of these seemingly conflicting requirements. 
Finally, this study demonstrates the value of employing computer science research and scalable audit strategies to effectively monitor and enforce GDPR compliance.

\section*{Acknowledgments}
We 
extend our sincere gratitude to Prof. Christoph Engel (Max Planck Institute for research on Collective Goods) and Prof. Cristoph Sorge (University of Saarland) for their valuable feedback. 
Additionally, Ingmar Weber is supported by funding from the Alexander von Humboldt Foundation and its founder, the Federal Ministry of Education and Research (Bundesministerium für Bildung und Forschung).
\bibliographystyle{plain}
\bibliography{main}
\appendices
\label{Sec:Appendix}
\begin{figure}
    \centering
    \begin{subfigure}[t]{0.45\columnwidth}
        \centering
        \includegraphics[width=\textwidth, height=3cm]{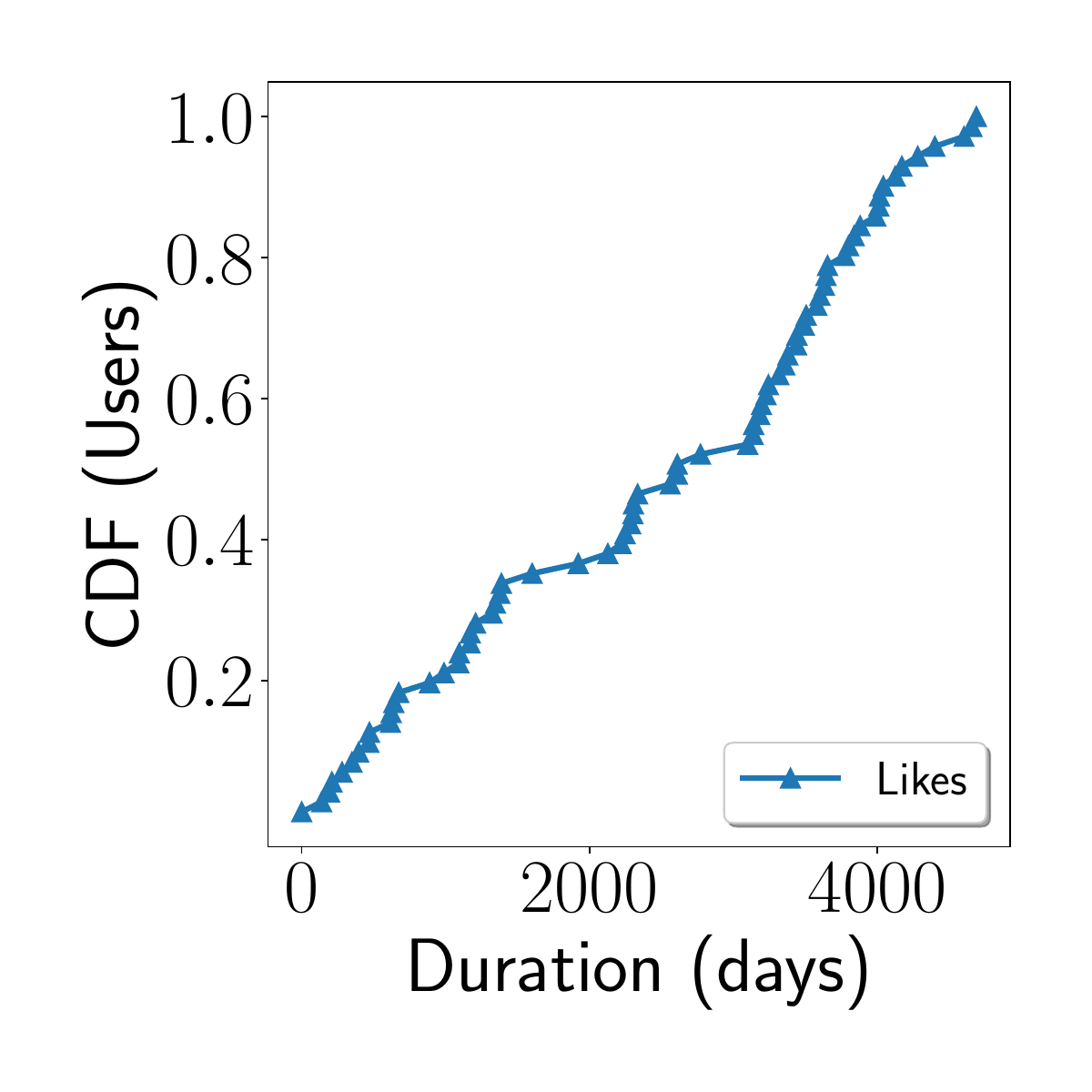}
         \caption{Instagram}
        \label{fig:instagram_cdf_like}
    \end{subfigure}
    \begin{subfigure}[t]{0.45\columnwidth}
        \centering
        \includegraphics[width=\textwidth, height=3cm]{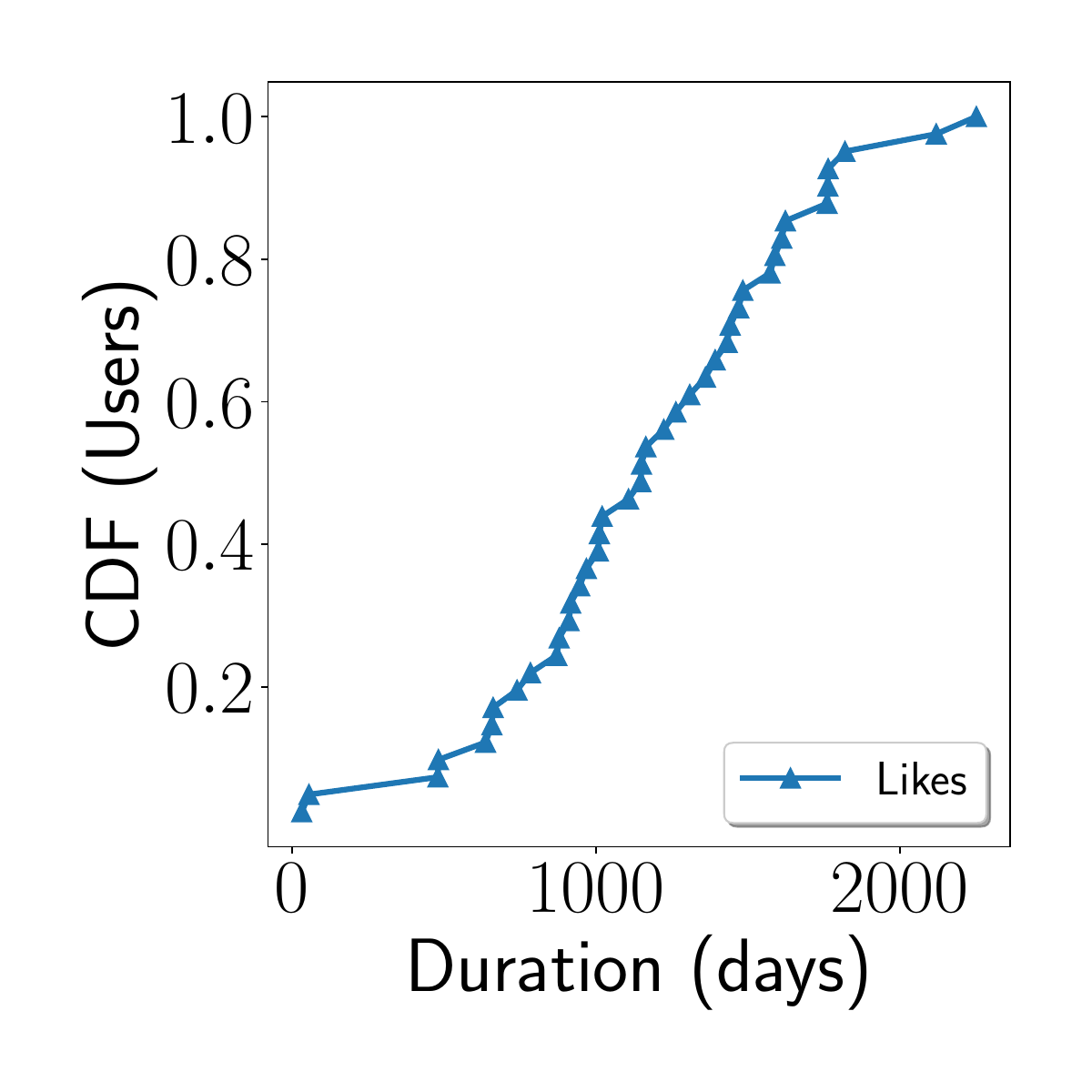}
         \caption{TikTok}
        \label{fig:tiktok_cdf_like}
    \end{subfigure}
    \hfill
    
    \caption{Plots illustrating the CDF of data duration provided to users for like activity.}
    

    \label{cdf_ike_overall}
\end{figure}

\begin{figure*}
    \centering
    \includegraphics[width=\textwidth, height=5cm]{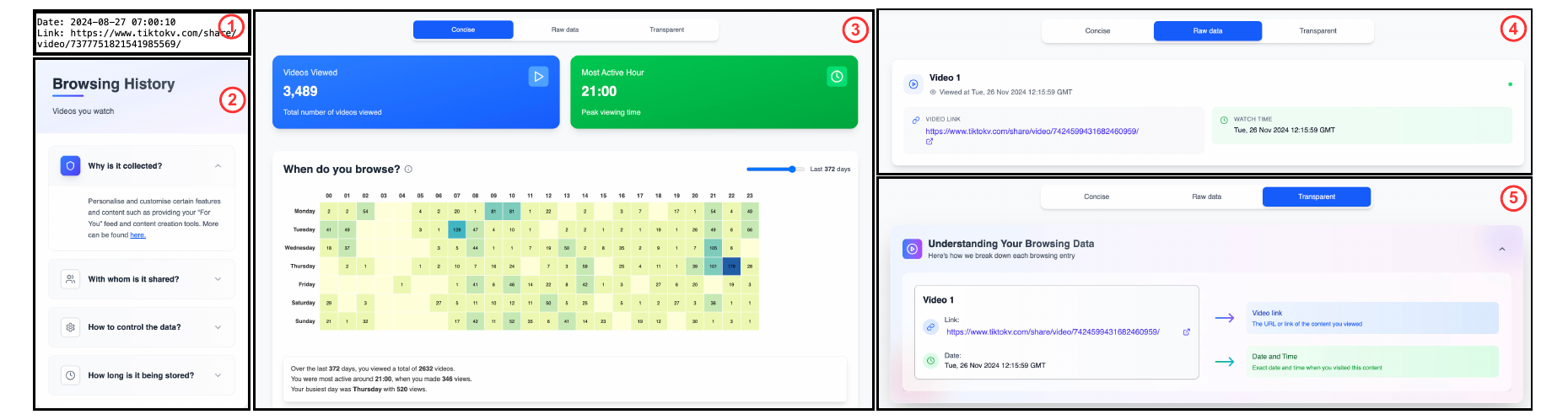}
    \caption{ \textcircled{1} shows browsing history in TikTok DDPs. \textcircled{2} illustrates how our approach addresses the transparency concerns raised by users. \textcircled{3}, \textcircled{4}, and \textcircled{5} depict our proposed concise, raw data, and transparent views respectively. 
    }
    \label{fig:recommendation_browse_tiktok}
\end{figure*}

\begin{figure*}
    \centering
    \includegraphics[width=\textwidth, height=5cm]{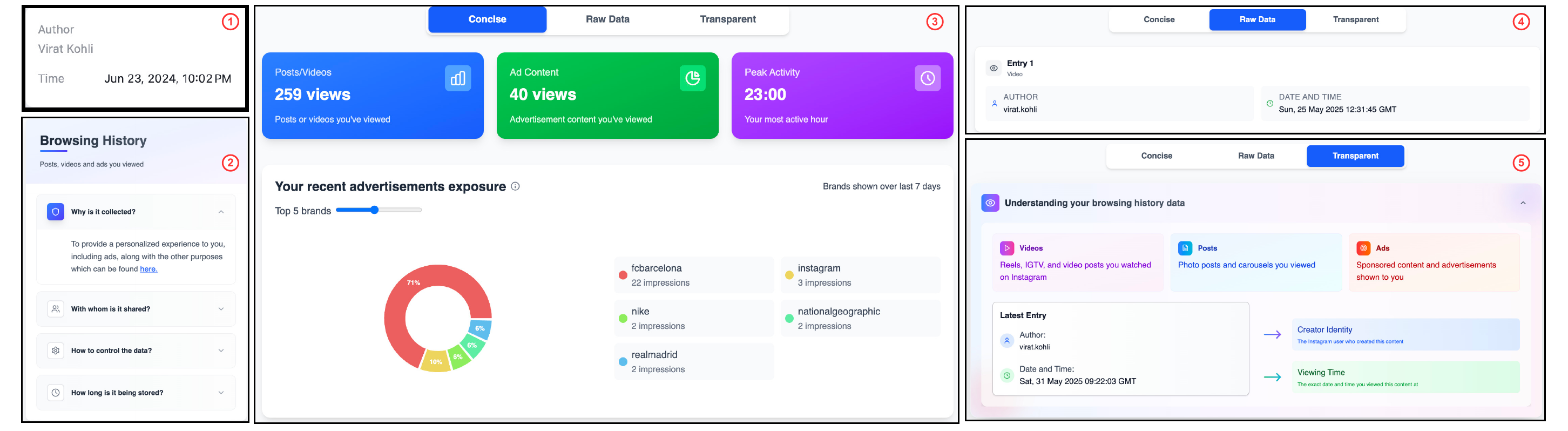}
    \caption{ \textcircled{1} shows watch history in Instagram DDPs. \textcircled{2} illustrates how our approach addresses the transparency concerns raised by users. \textcircled{3}, \textcircled{4}, and \textcircled{5} depict our proposed concise, raw data, and transparent views respectively. 
    }
    \label{fig:recommendation_browse_instagram}
\end{figure*}

\begin{table*}
    \centering
   \footnotesize
   \begin{tabular}{@{}lcrr|rr|rr|rr|@{}}
\toprule
\textbf{}                & \textbf{}     & \multicolumn{2}{c}{\textbf{\Cref{Sec: Reliability}}} & \multicolumn{2}{c}{\textbf{\Cref{Sec: Comprehensibility}}}  & \multicolumn{2}{c}{\textbf{\Cref{Sec: RecommendationRepresentation}}} & \multicolumn{2}{c}{\textbf{\Cref{Sec: RecommendataionComprehensibility}}} \\ \toprule
\textbf{Attribute}                & \textbf{Type}     & \multicolumn{1}{c}{\textbf{Count}} & \multicolumn{1}{c}{\textbf{Percentage}}  & \multicolumn{1}{c}{\textbf{Count}} & \multicolumn{1}{c}{\textbf{Percentage}} & \multicolumn{1}{c}{\textbf{Count}} & \multicolumn{1}{c}{\textbf{Percentage}}  & \multicolumn{1}{c}{\textbf{Count}} & \multicolumn{1}{c}{\textbf{Percentage}}\\ \midrule
\multirow{4}{*}{\textbf{Gender}}  
& Male   & 37 & 46.2  & 228 & 57.0 & 100 & 50.0 & 58 & 48.3 \\
& Female & 34 & 42.5 & 161 & 40.3  & 99 & 49.5 & 59 & 49.2 \\
& Other  & 0 & 0.0  & 9 & 2.3 & 0 & 0 & 1 & 0.8 \\
& Prefer not to say & 9 & 11.2 & 2 & 0.5 & 1 & 0.5  & 2 & 1.7  \\ \midrule
\multirow{4}{*}{\textbf{Age}} 
& 18-25  & 25 & 31.2 & 95 & 23.7 & 54 & 27.0 & 34 & 28.3 \\
& 25-30 & 18 & 22.5 & 104 & 26.0  & 57 & 28.5 & 36 & 30.0  \\
& 30-40  & 19 & 23.8  & 114 & 28.5 & 57 & 28.5 & 32 & 26.7 \\
& 40+   & 9 & 11.2 & 87 & 21.8 & 32 & 16.0 & 18 & 15.0 \\ 
& Prefer not to say & 9 & 11.2 & 0 & 0.0 & 0 & 0.0 & 0 & 0.0\\
\midrule
\multirow{4}{*}{\textbf{Country}} & 
France  & 10 & 11.2 & 100 & 25.0  & 50 & 25.0 & 30 & 25.0 \\
 & Germany & 10 & 11.2 & 100  & 25.0 & 50 & 25.0 & 30 & 25.0 \\
& Spain & 10 & 11.2 & 100  & 25.0 & 50 & 25.0 & 30 & 25.0 \\
& Italy & 10 & 11.2 & 100 & 25.0 & 50 & 25.0 & 30 & 25.0 \\ 
& Brazil & 10 & 11.2 & 0 & 0.0 & 0 & 0.0 & 0 & 0.0\\
& USA & 10 & 11.2 & 0 & 0.0 & 0 & 0.0 & 0 & 0.0 \\
& UK & 10 & 11.2 & 0 & 0.0 & 0 & 0.0 & 0 & 0.0\\
& India & 10 & 11.2 & 0 & 0.0 & 0 & 0.0 & 0 & 0.0\\
\bottomrule
\end{tabular}
    \caption{Distribution of participants recruited in \Cref{Sec: Reliability}, \Cref{Sec: Comprehensibility},  \Cref{Sec: RecommendationRepresentation} and \Cref{Sec: RecommendataionComprehensibility}  based on their self-reported gender, age, and country of residence. }
    \label{Tab: Demographics}
\end{table*}

\label{appendix:reliability}

\section{Reliability of DDPs}
\Cref{cdf_ike_overall} represents the plots that show the CDF of data duration provided to users for the like history for the two platforms - Instagram and TikTok. We do not observe any clusters, suggesting that these platforms provide the like history more consistently to the end users.

\section{User demographics for different surveys}
\label{appendix:survey}

\Cref{Tab: Demographics} presents the distribution of the participants recruited  in \Cref{Sec: Reliability}, \Cref{Sec: Comprehensibility},  \Cref{Sec: RecommendationRepresentation} and \Cref{Sec: RecommendataionComprehensibility}. 

\section{Improving DDPs comprehensibility}
\label{appendix:improve_comprehensibility}
\Cref{fig:recommendation_browse_tiktok} and \Cref{fig:recommendation_browse_instagram} are the partial screenshots of the dashboard for browsing history on TikTok and Instagram as proposed in our extension \textit{Know Your Data}.
\end{document}